\documentclass[twocolumn,showpacs,amsmath,amssymb,superscriptaddress]{revtex4-1}
\usepackage{dcolumn}
\usepackage{color}
\usepackage{graphicx}
\usepackage{amsmath}
\usepackage{multirow}
\usepackage{bm}
\usepackage{url}

\newcommand\+{\dagger}

\begin{document}

\title{Spectroscopy of odd-odd nuclei within the interacting boson-fermion-fermion model based on the Gogny energy density functional}

\author{K.~Nomura}
\affiliation{Physics Department, Faculty of Science, University of
Zagreb, HR-10000 Zagreb, Croatia}
\affiliation{Advanced Science Research Center, Japan Atomic Energy
Agency, Tokai, 319-1195 Ibaraki, Japan}

\author{R.~Rodr\'iguez-Guzm\'an}
\affiliation{Physics Department, Kuwait University, 13060 Kuwait, Kuwait}

\author{L.~M.~Robledo}
\affiliation{Departamento de F\'\i sica Te\'orica, Universidad
Aut\'onoma de Madrid, E-28049 Madrid, Spain}

\affiliation{Center for Computational Simulation,
Universidad Polit\'ecnica de Madrid,
Campus de Montegancedo, Boadilla del Monte, 28660-Madrid, Spain
}

\date{\today}

\begin{abstract}
We present a method to calculate spectroscopic properties of odd-odd 
nuclei within the framework of the Interacting Boson-Fermion-Fermion 
Model based on the Gogny energy density functional. The 
$(\beta,\gamma)$-deformation energy surface of the even-even 
(boson-)core nucleus, spherical single-particle energies and occupation 
probabilities of the odd neutron and odd proton, are provided by the 
constrained self-consistent mean-field calculation within the 
Hartree-Fock-Bogoliubov method with the Gogny-D1M functional. These 
quantities are used as a microscopic input to fix most of the 
parameters of the IBFFM Hamiltonian. Only a few coupling constants for 
the boson-fermion Hamiltonian and the residual neutron-proton 
interaction are specifically adjusted to reproduce experimental 
low-energy spectra in odd-mass and odd-odd nuclei, respectively. In 
this way,  the number of free parameters involved in the IBFFM 
framework is reduced significantly. The method is successfully applied 
to the description of the low-energy spectra and  electromagnetic 
transition rates in the odd-odd $^{194,196,198}$Au nuclei. 
\end{abstract}


\maketitle


\section{Introduction}


The unified theoretical description of low-lying states in even-even, 
odd-mass, and odd-odd nuclei is one of the major goals of nuclear 
structure. In even-even systems at low energy, nucleons are coupled 
pairwise and the type of couplings determines the low-lying collective 
structure of vibrational and rotational states. The microscopic 
description of low-lying collective states in even-even systems has 
been extensively pursued with numerous theoretical methods 
\cite{BM,RS,IBM,CasBook,bender2003,caurier2005}. However, the 
description of odd-mass and odd-odd nuclei is more cumbersome, due to 
the fact that in those systems both collective and single-particle 
motions have to be treated on the same footing \cite{BM,bohr1953}.

The Interacting Boson Model (IBM) \cite{IBM} has been remarkably successful in
the phenomenological study of low-lying structures in medium-mass and heavy even-even nuclei. 
In its simplest version, the building blocks of the IBM are the monopole
$s$ and quadrupole $d$ bosons, which represent the collective pairs of
valence nucleons coupled to spin and parity $0^+$ and $2^+$, respectively \cite{IBM,OAI}.
The microscopic foundation of the IBM starting from the nucleonic degrees
of freedom has been extensively pursued in the literature
\cite{OAIT,OAI,otsuka1984,mizusaki1997,nomura2008,nomura2011rot}. In
particular, a systematic
method of deriving the IBM Hamiltonian from microscopic input has been developed in
\cite{nomura2008}. 
In this approach, the deformation energy surface that is obtained from
the self-consistent mean-field (SCMF) calculation based on a given energy
density functional (EDF) is mapped onto the expectation value of the IBM
Hamiltonian in the boson coherent state \cite{ginocchio1980}. This
procedure completely determines the strength parameters of the IBM Hamiltonian. 
Since the EDF framework allows for a global mean-field description of
intrinsic properties of nuclei over the entire  Segr\'e's
chart, it has become possible to determine in a unified way the 
parameters of the IBM Hamiltonian basically
for any arbitrary nucleus.

The method mentioned above has been recently 
extended to odd-mass systems \cite{nomura2016odd} by considering the coupling between bosonic
(collective) degrees of freedom and an unpaired nucleon within the
framework of the Interacting Boson-Fermion Model (IBFM) \cite{IBFM}. 
In this extension, the even-even core (IBM) Hamiltonian, the single-particle energies and occupation
probabilities of the odd particle, which are building blocks of the IBFM
Hamiltonian, have been completely determined based on the output of a SCMF calculation. 
Even though a few strength parameters for the particle-boson coupling
are treated as free
parameters, the method allows for
an accurate, systematic, and computationally feasible description of 
various low-energy properties of odd-mass medium-mass and heavy nuclei: 
e.g., signatures of shape phase transitions 
\cite{nomura2016qpt,nomura2017odd-1,nomura2017odd-3,nomura2018pt}, octupole
correlations in neutron-rich odd-mass Ba 
isotopes \cite{nomura2018oct}, and the structure of neutron-rich
odd-mass Kr isotopes \cite{nomura2018kr}.

In this work, we extend these studies to odd-odd nuclei by using
the Interacting Boson-Fermion-Fermion Model (IBFFM)
\cite{brant1984,IBFM}. 
The IBFFM is an extension of the IBFM that considers odd-odd nuclei as a 
system composed of an IBM core plus an unpaired neutron and an unpaired proton.
The IBM-core and particle-boson coupling Hamiltonians are determined in
way similar to that employed for odd-mass nuclei \cite{nomura2016odd}. 
The only additional parameters are the coefficients of the residual
neutron-proton interaction. They are determined to reasonably reproduce
the experimental data for the low-lying spectra of the considered odd-odd
nuclei. 
The microscopic input used to determine part of the IBFFM Hamiltonian is obtained by
constrained SCMF calculations within the Hartree-Fock-Bogoliubov (HFB) method
based on the Gogny D1M EDF \cite{D1M}.  
The two most relevant parametrizations of the finite range Gogny force, namely
D1S \cite{D1S} and D1M \cite{D1M} have proven along the years to
provide a reliable description of many collective phenomena all over
the periodic table (see \cite{Delaroche2010,Robledo2011} for some examples).
Our choice of D1M is based solely on its better performance to describe
binding energies.

As an application of the proposed methodology, we specifically study the
properties of the odd-odd $^{194,196,198}$Au nuclei.  
Their low-lying structures are described by unpaired neutron and proton
holes coupled with the even-even core nuclei $^{196,198,200}$Hg. 
The IBM parameters for the even-even cores were already obtained in Ref.~\cite{nomura2013hg}
as part of a comprehensive study of  shape coexistence
and low-lying structures in the entire Hg isotopic chain within the
configuration-mixing IBM method based on the Gogny-D1M EDF. The results obtained
suggest that the nuclei $^{196,198,200}$Hg have weakly oblate deformed to
nearly spherical ground-state shapes. 
For the neighbouring odd-N nuclei $^{195,197,199}$Hg and odd-Z nuclei
$^{195,197,199}$Au, there are plenty of experimental data to determine
the boson-fermion strength parameters. 
In addition, the odd-odd Au nuclei in this mass region have previously
been extensively studied within the IBFFM framework: e.g., by means of 
numerical studies \cite{lopac1986,blasi1987}, or by 
pure-algebraic approaches \cite{vanisacker1985,barea2005,thomas2014} in the
context of nuclear supersymmetry \cite{iachello1980}. Those
results will be a good reference to compare with our less phenomenological
results.

On the other hand, it is worth to mention that microscopic nuclear
structure models are also  applied in the
spectroscopic studies of odd-mass and/or odd-odd nuclei with the Gogny force. 
As an example, let us mention the studies of 
various low-energy properties of
odd-mass systems at the mean-field level using full blocking 
\cite{PhysRevC.86.064313} or the equal filling approximation \cite{rayner10odd-1,rayner10odd-2,rayner10odd-3}.
To our knowledge there is only one study \cite{Robledo2014} of odd-odd 
nuclei focused on the ability to reproduce the empirical Gallagher-Moszkowski (GM) rule.
As shown in this reference, the GM rule is not fulfilled by the Gogny force
and the failure is traced back to the lack of additional proton-neutron
interaction terms in the interaction. This difficulty and the inability
of any effective interaction to reach spectroscopic accuracy for the 
spectra of odd nuclei \cite{Dobaczewski2015} point to the necessity to add extra terms with
extra parameters that can be fitted locally to improve the quality of 
the description of odd and odd-odd nuclei. This is achieved in our model
through the set of extra terms added with parameters not fixed by the EDF input.
Another source of difficulties hampering to reach  spectroscopic accuracy in the description of odd nuclei 
with EDFs is the impact of dynamical 
correlations as those coming from symmetry restoration \cite{RS}.  In the
last few years it has been possible to include time-reversal symmetry and blocking effects
along with angular momentum and particle number projection \cite{bally2014}
but the complexity of the problem prevents its use beyond very light 
systems like $^{24}$Mg \cite{bally2014,borrajo2016}.

This paper is organized as follows: In Sec.~\ref{sec:model} we describe
the procedure to construct the IBFFM Hamiltonian based on the SCMF
calculation. 
In Sec.~\ref{sec:results}, the spectroscopic properties of 
the even-even Hg nuclei are briefly reviewed. In the same section, the
results for low-energy spectra in the odd-N Hg and the odd-Z Au isotopes
are discussed, followed by the results of the spectroscopic calculations
for the odd-odd Au nuclei. 
Finally, a short summary and concluding remarks are given in
Sec.~\ref{sec:summary}.


\section{Theoretical framework\label{sec:model}}


\subsection{Hamiltonian}

In this work we use the version of the IBFFM that distinguishes between
neutron and proton degrees of freedom (denoted hereafter as IBFFM-2). 
The IBFFM-2 Hamiltonian is expressed as: 
\begin{equation}
\label{eq:ham}
 \hat H_\text{} = \hat H_\text{B} + \hat H_\text{F}^\nu + \hat
 H_\text{F}^\pi + \hat
  H_\text{BF}^\nu + H_\text{BF}^\pi + \hat V_\text{res}. 
\end{equation}
The first term in Eq.~(\ref{eq:ham}) is the neutron-proton IBM (IBM-2)
Hamiltonian \cite{OAI}  that describes the even-even core nuclei
$^{196,198,200}$Hg. The second and third terms represent the
Hamiltonian for an odd neutron and an odd proton, respectively. 
The fourth and fifth terms correspond to the interaction Hamiltonians describing the couplings
of the odd
neutron and of the odd proton to the IBM-2 core, respectively.
The last term in Eq.~(\ref{eq:ham}) is the residual interaction between
the odd neutron and odd proton. 

For the boson-core Hamiltonian $\hat H_\text{B}$ the standard IBM-2 Hamiltonian is adopted: 
\begin{equation}
\label{eq:ibm2}
 \hat H_{\text{B}} = \epsilon(\hat n_{d_\nu} + \hat n_{d_\pi})+\kappa\hat
  Q_{\nu}\cdot\hat Q_{\pi} + \kappa'\hat L\cdot\hat L,
\end{equation}
where $\hat n_{d_\rho}=d^\dagger_\rho\cdot\tilde d_{\rho}$ ($\rho=\nu,\pi$) is the
$d$-boson number operator, $\hat Q_\rho=d_\rho^\dagger s_\rho +
s_\rho^\dagger\tilde d_\rho^\dagger + \chi_\rho(d^\dagger_\rho\times\tilde
d_\rho)^{(2)}$ is the quadrupole operator, and $\hat L=\hat L_\nu + \hat
L_\pi$ is the angular momentum operator with $\hat
L_{\rho}=\sqrt{10}(d_\rho^\dagger\times\tilde d_\rho)^{(1)}$. The different parameters
of the Hamiltonian are denoted by 
$\epsilon$, $\kappa$, $\chi_\nu$, $\chi_\pi$, and $\kappa'$. 
The doubly-magic nucleus $^{208}$Pb is taken as the inert 
core for the boson space. The numbers of neutron $N_{\nu}$ and proton
$N_{\pi}$ bosons equal the number of 
neutron-hole and proton-hole pairs, respectively. As a consequence,  $N_\pi=1$ and $N_{\nu}=5$, 4, and
3 for the $^{196,198,200}$Hg nuclei, respectively. 

The Hamiltonian for the odd nucleon reads: 
\begin{equation}
 \hat H_\text{F}^\rho = -\sum_{j_\rho}\epsilon_{j_\rho}\sqrt{2j_\rho+1}
  (a_{j_\rho}^\dagger\times\tilde a_{j_\rho})^{(0)}
\end{equation}
with $\epsilon_{j_\rho}$ being the single-particle energy of the odd
nucleon. 
$j_\nu$ ($j_\pi$) stands for the angular momentum of the odd neutron
(proton).
$a_{j_\rho}^{(\dagger)}$ represents the fermion annihilation (creation)
operator, and $\tilde a_{j_\rho}$ is defined as $\tilde a_{jm}=(-1)^{j-m}a_{j-m}$.
For the fermion valence space, we consider the full neutron major
shell $N=82-126$, i.e., $3p_{1/2}$, $3p_{3/2}$, $2f_{5/2}$, $2f_{7/2}$, 
$1h_{9/2}$, and $1i_{13/2}$ orbitals, and the full proton major shell
$Z=50-82$, i.e., $3s_{1/2}$, $2d_{3/2}$, $2d_{5/2}$, $1g_{7/2}$, and
$1h_{11/2}$ orbitals.

For the boson-fermion interaction term $\hat H_{\rm BF}^\rho$ in
Eq.~(\ref{eq:ham}), we use the following form: 
\begin{equation}
\label{eq:ham-bf}
 \hat H_\text{BF}^\rho = \Gamma_\rho\hat Q_{\rho'}\cdot\hat q_{\rho} 
+
  \Lambda_\rho\hat V_{\rho'\rho} + A_\rho\hat n_{d_{\rho}}\hat n_{\rho}
\end{equation}
where $\rho'\neq\rho$, and the first, second, and third terms are the
quadrupole dynamical, exchange, and monopole terms, respectively. The parameters
of the interaction Hamiltonian are denoted by  
$\Gamma_\rho$, $\Lambda_\rho$, and $A_{\rho}$. 
As in the previous studies
\cite{scholten1985,arias1986}, we assume that both the dynamical and exchange terms 
are dominated by the interaction between unlike particles (i.e., between
the odd neutron and proton bosons and between the odd proton and neutron
bosons). We also assume that for the monopole term the interaction between
like-particles (i.e., between the odd neutron and neutron bosons and between
the odd proton and proton bosons) plays a dominant role. 
In Eq.~(\ref{eq:ham-bf}) $\hat Q_\rho$ is the same bosonic quadrupole operator as 
in the IBM-2 Hamiltonian in Eq.~(\ref{eq:ibm2}). The fermionic quadrupole 
operator $\hat q_\rho$ reads: 
\begin{equation}
\hat q_\rho=\sum_{j_\rho j'_\rho}\gamma_{j_\rho j'_\rho}(a^\+_{j_\rho}\times\tilde
a_{j'_\rho})^{(2)},
\end{equation} 
where $\gamma_{j_\rho
j'_\rho}=(u_{j_\rho}u_{j'_\rho}-v_{j_\rho}v_{j'_\rho})Q_{j_\rho
j'_\rho}$ and  $Q_{j_\rho j'_\rho}=\langle
l\frac{1}{2}j_{\rho}||Y^{(2)}||l'\frac{1}{2}j'_{\rho}\rangle$ represents
the matrix element of the fermionic 
quadrupole operator in the considered single-particle basis.
The exchange term $\hat V_{\rho'\rho}$ in Eq.~(\ref{eq:ham-bf}) reads: 
\begin{align}
\label{eq:Rayner-new-label}
 \hat V_{\rho'\rho} =& -(s_{\rho'}^\+\tilde d_{\rho'})^{(2)}
\cdot
\Bigg\{
\sum_{j_{\rho}j'_{\rho}j''_{\rho}}
\sqrt{\frac{10}{N_\rho(2j_{\rho}+1)}}\beta_{j_{\rho}j'_{\rho}}\beta_{j''_{\rho}j_{\rho}} \nonumber \\
&:((d_{\rho}^\+\times\tilde a_{j''_\rho})^{(j_\rho)}\times
(a_{j'_\rho}^\+\times\tilde s_\rho)^{(j'_\rho)})^{(2)}:
\Bigg\} + (H.c.), \nonumber \\
\end{align}
with $\beta_{j_{\rho}j'_{\rho}}=(u_{j_{\rho}}v_{j'_{\rho}}+v_{j_{\rho}}u_{j'_{\rho}})Q_{j_{\rho}j'_{\rho}}$.
In the second line of the above equation the notation $:(\cdots):$ indicates normal ordering. 
In the monopole interactions, the number operator for the odd fermion is
expressed as $\hat
n_{\rho}=\sum_{j_{\rho}}(-\sqrt{2j_{\rho}+1})(a^\+_{j_\rho}\times\tilde 
a_{j_\rho})^{(0)}$.

In previous IBFFM
calculations \cite{brant2006,yoshida2013}, the residual interaction $\hat V_\text{res}$ in Eq.~(\ref{eq:ham}) contained a quadrupole-quadrupole,
delta, spin-spin-delta, spin-spin, and tensor interaction. 
However, we find that only the delta and spin-spin-delta terms are 
enough to provide a good description of the low-lying states in the odd-odd nuclei considered
here. Therefore, the residual interaction used here reads: 
\begin{equation}
 \hat V_{\text{res}}=4\pi\delta({\bf r_\nu}-{\bf
  r_\pi})(u_0+u_1{\bf\sigma_\nu}\cdot{\bf\sigma_\pi}), 
\end{equation}
with $u_0$ and $u_1$ the parameters. 
Furthermore, the matrix element of the residual interaction 
$\hat V_\text{res}$, denoted by $V_\text{res}'$, can be expressed as
\cite{yoshida2013}: 
\begin{align}
\label{eq:vres}
V_\text{res}'
&= (u_{j_\nu'} u_{j_\pi'} u_{j_\nu} u_{j_\nu} + v_{j_\nu'} v_{j_\pi'} v_{j_\nu} v_{j_\nu})
V^{J}_{j_\nu' j_\pi' j_\nu j_\pi}
\nonumber \\
& {} - (u_{j_\nu'}v_{j_\pi'}u_{j_\nu}v_{j_\pi} +
 v_{j_\nu'}u_{j_\pi'}v_{j_\nu}u_{j_\pi}) \nonumber \\
&\times \sum_{J'} (2J'+1)
\left\{ \begin{array}{ccc} {j_\nu'} & {j_\pi} & J' \\ {j_\nu} & {j_\pi'} & J
\end{array} \right\} 
V^{J'}_{j_\nu'j_\pi j_\nu j_\pi'}, 
\end{align}
where
\begin{equation}
V^{J}_{j_\nu'j_\pi'j_\nu j_\pi} = \langle j_\nu'j_\pi';J|\hat
 V_\text{res}|j_\nu j_\pi;J\rangle
\end{equation}
is the matrix element between the neutron-proton pairs, and $J$ stands
for the total angular momentum of the neutron-proton pair. 
The bracket in Eq.~(\ref{eq:vres}) stands for the Racah coefficient. 
Also in Eq.~(\ref{eq:vres}) the terms resulting from contractions are ignored as in Ref.~\cite{morrison1981}.
A similar residual neutron-proton interaction is used in the
two-quasiparticle-rotor-model calculation in Ref.~\cite{bark1997}.

\subsection{Procedure to build the IBFFM-2 Hamiltonian}

The ingredients of the  IBFFM-2 Hamiltonian $\hat H$ in
Eq.~(\ref{eq:ham}) are determined with the following procedure. 
\begin{enumerate}
 \item Firstly, the IBM-2 Hamiltonian is determined by using the methods of
       Refs.~\cite{nomura2008,nomura2011rot}: the $(\beta,\gamma)$-deformation
       energy surface obtained from the constrained Gogny-D1M HFB calculation is mapped
       onto the expectation value of the IBM-2 Hamiltonian in the boson
       coherent state \cite{ginocchio1980}. This procedure
       completely determines the parameters $\epsilon$, $\kappa$,
       $\chi_\nu$, and $\chi_\pi$ in the IBM-2 Hamiltonian. Only the
       strength parameter $\kappa'$ for the $\hat L\cdot\hat L$ term is
       determined separately from the other parameters, by adjusting the cranking moment of inertia in
       the boson intrinsic state to the corresponding Thouless-Valatin \cite{TV}
       moment of inertia obtained by the Gogny-HFB SCMF calculation at
       the equilibrium mean-field minimum \cite{nomura2011rot}. 

\item Second, the strength parameters for the boson-fermion coupling
      Hamiltonians $\hat H_\text{BF}^\nu$ and $\hat H_\text{BF}^\pi$ for
      the odd-N Hg and odd-Z Au nuclei, respectively, is
      determined by using the procedure of
      \cite{nomura2016odd}: Single-particle energies and 
      occupation probabilities of the odd nucleon are provided by the Gogny-HFB
      calculation constrained to zero deformation (see,
      Ref.~\cite{nomura2017odd-2}, for details); Optimal values of the
      parameters $\Gamma_\nu$, $\Lambda_\nu$, and $A_\nu$ 
      ($\Gamma_\pi$, $\Lambda_\pi$, and $A_\pi$), are chosen,
      separately for positive and negative parity, so as to reproduce
      the experimental low-energy levels of each of the considered odd-N Hg (odd-Z Au) nuclei. 

\item By following previous IBFFM calculations
      \cite{brant1984,brant2006,yoshida2013}, the same strength parameters $\Gamma_\nu$, $\Lambda_\nu$, and $A_\nu$
      ($\Gamma_\pi$, $\Lambda_\pi$, and $A_\pi$) as those obtained for
      the odd-N Hg (odd-Z Au) nuclei in
      the previous step, are used for the odd-odd nuclei. The
      single-particle energies and occupation probabilities are, 
      however, newly calculated for the odd-odd systems. 

\item Finally, the parameters in the residual 
      interaction $\hat V_{\text{res}}$, i.e., $u_0$ and $u_1$, are
      determined so as to reasonably reproduce the low-lying
      spectra in the studied odd-odd nuclei. The fixed values $u_0=-0.3$ MeV
      and $u_1=-0.033$ MeV for positive parity, 
      and $u_0=-0.3$ MeV and $u_1=0.0$ MeV for negative-parity
      states, are adopted. The ratio, $u_0/u_1\approx 9$, was also
      considered in \cite{bark1997}. 
\end{enumerate}

The values of the IBM-2 parameters employed in the present work are
shown in Table~\ref{tab:ibm2para}. They are exactly the same as those
used in Ref.~\cite{nomura2013hg}. 
The fitted strength parameters for the Hamiltonian $\hat H_\text{BF}^\nu$
($\hat H_\text{BF}^\pi$), i.e., $\Gamma_\nu$, $\Lambda_\nu$, and
$A_\nu$ ($\Gamma_\pi$, $\Lambda_\pi$, and $A_\pi$) are shown in Table~\ref{tab:bf-oddn}
(Table~\ref{tab:bf-oddz}). 
The fixed value $\Gamma_\rho=0.8$ MeV is used for the strength
parameter for the quadrupole dynamical term for
all the odd-mass and odd-odd nuclei and for both parities. 
Other parameters do not differ too much between neighbouring isotopes. 
Tables~\ref{tab:vsq-oddn}, \ref{tab:vsq-oddz}, and \ref{tab:vsq-dodd} summarize the
single-particle energies and occupation probabilities obtained from the
Gogny-HFB SCMF calculations for the studied odd-N Hg, odd-Z Au, and
odd-odd Au isotopes, respectively. 
We note that the single-particle energies and occupation probabilities for the odd-N
Hg (Table~\ref{tab:vsq-oddn}) and odd-Z Au (Table~\ref{tab:vsq-oddz})
nuclei are almost identical to those computed for the odd-odd Au nuclei (see,
Table~\ref{tab:vsq-dodd}).

\begin{table}[htb!]
 \begin{center}
\caption{\label{tab:ibm2para} The adopted parameters of the IBM-2
  Hamiltonian $\hat H_\text{B}$ in Eq.~(\ref{eq:ibm2}). They are taken from Ref.~\cite{nomura2013hg}.}
  \begin{tabular}{cccccc}
\hline\hline
   & $\epsilon$ (MeV) & $\kappa$ (MeV) & $\chi_\nu$ & $\chi_\pi$ &
   $\kappa'$ (MeV) \\
\hline
$^{196}$Hg & 0.710 & -0.517 & 0.836 & 0.613 & 0.0041 \\
$^{198}$Hg & 0.675 & -0.470 & 1.333 & 0.166 & 0.0043 \\
$^{196}$Hg & 0.636 & -0.328 & 0.891 & 0.684 & 0.0018 \\
\hline\hline
  \end{tabular}
 \end{center}
\end{table}


\begin{table}[htb!]
\caption{\label{tab:bf-oddn} Strength parameters of the Hamiltonian
 $\hat H_\text{BF}^\nu$ (in MeV ) employed for the
 odd-N nuclei $^{195,197,199}$Hg and odd-odd nuclei $^{194,196,198}$Au.}
 \begin{center}
  \begin{tabular}{ccccccc}
\hline\hline
   & $\Gamma_\nu^+$ & $\Lambda_\nu^+$ & $A_\nu^+$ & $\Gamma_\nu^-$ & $\Lambda_\nu^-$ & $A_\nu^-$\\
\hline
$^{194}$Au,$^{195}$Hg & 0.80 & 0.0 & -0.10 & 0.80 & 2.00 & -0.80 \\
$^{196}$Au,$^{197}$Hg & 0.80 & 0.0 & 0.0 & 0.80 & 1.50 & -0.40 \\
$^{198}$Au,$^{199}$Hg & 0.80 & 0.0 & -0.20 & 0.80 & 1.20 & -0.35 \\
\hline\hline
  \end{tabular}
 \end{center}
\end{table}

\begin{table}[htb!]
\caption{\label{tab:bf-oddz} Strength parameters of the Hamiltonian
 $\hat H_\text{BF}^\pi$ (in MeV ) employed for the
 odd-Z nuclei $^{195,197,199}$Au and odd-odd nuclei $^{194,196,198}$Au.}
 \begin{center}
  \begin{tabular}{ccccccc}
\hline\hline
   & $\Gamma_\pi^+$ & $\Lambda_\pi^+$ & $A_\pi^+$ & $\Gamma_\pi^-$ & $\Lambda_\pi^-$ & $A_\pi^-$\\
\hline
$^{194,195}$Au & 0.80 & 1.50 & 0.0 & 0.80 & 1.50 & -0.80 \\
$^{196,197}$Au & 0.80 & 1.60 & 0.0 & 0.80 & 0.00 & 0.0 \\
$^{198,199}$Au & 0.80 & 2.40 & 0.0 & 0.80 & 0.00 & 0.0 \\
\hline\hline
  \end{tabular}
 \end{center}
\end{table}


\begin{table}[htb!]
\caption{\label{tab:vsq-oddn} Neutron single-particle energies $\epsilon_{j_{\nu}}$ (in MeV )
 and occupation probabilities $v^2_{j_\pi}$ used in the present study for the odd-N nuclei $^{195,197,199}$Hg.}
 \begin{center}
  \begin{tabular}{cccccccc}
\hline\hline
   & & $3p_{1/2}$ & $3p_{3/2}$ & $2f_{5/2}$ & $2f_{7/2}$ & $1h_{9/2}$ & $1i_{13/2}$\\
\hline
$^{195}$Hg & $\epsilon_{j_\nu}$ & 0.000 & 0.921 & 1.033 & 3.819 & 4.283 & 1.537 \\
           & $v^2_{j_\nu}$ & 0.248 & 0.515 & 0.554 & 0.944 & 0.951 & 0.702 \\
\hline
$^{197}$Hg & $\epsilon_{j_\nu}$ & 0.000 & 0.937 & 1.056 & 3.846 & 4.366 & 1.570 \\
           & $v^2_{j_\nu}$ & 0.289 & 0.590 & 0.631 & 0.956 & 0.962 & 0.769 \\
\hline
$^{199}$Hg & $\epsilon_{j_\nu}$ & 0.000 & 0.957 & 1.078 & 3.877 & 4.449 & 1.605 \\
           & $v^2_{j_\nu}$ & 0.338 & 0.670 & 0.713 & 0.967 & 0.973 & 0.834 \\
\hline\hline
  \end{tabular}
 \end{center}
\end{table}

\begin{table}[htb!]
\caption{\label{tab:vsq-oddz} Proton single-particle energies
 $\epsilon_{j_{\pi}}$ and occupation probabilities $v^2_{j_\pi}$ used in
 the present study for the
 odd-Z nuclei $^{195,197,199}$Au.}
 \begin{center}
  \begin{tabular}{ccccccc}
\hline\hline
   & & $3s_{1/2}$ & $2d_{3/2}$ & $2d_{5/2}$ & $1g_{7/2}$ & $1h_{11/2}$\\
\hline
$^{195}$Au & $\epsilon_{j_\pi}$ & 0.000 & 0.907 & 2.624 & 5.163 & 0.840 \\
           & $v^2_{j_\pi}$ & 0.617 & 0.870 & 0.968 & 0.989 & 0.864 \\
\hline
$^{197}$Au & $\epsilon_{j_\pi}$ & 0.00 & 0.888 & 2.592 & 5.153 & 0.834 \\
           & $v^2_{j_\pi}$ & 0.619 & 0.869 & 0.968 & 0.989 & 0.865 \\
\hline
$^{199}$Au & $\epsilon_{j_\pi}$ & 0.000 & 0.865 & 2.559 & 5.133 & 0.817 \\
           & $v^2_{j_\pi}$ & 0.624 & 0.867 & 0.967 & 0.989 & 0.864 \\
\hline\hline
  \end{tabular}
 \end{center}
\end{table}


\begin{table*}[htb!]
\caption{\label{tab:vsq-dodd} Neutron and proton single-particle
 energies (in MeV) and occupation probabilities used in the
 present study for
 the odd-odd nuclei $^{194,196,198}$Au.}
 \begin{center}
  \begin{tabular}{ccccccccccccccc}
\hline\hline
   & & $3p_{1/2}$ & $3p_{3/2}$ & $2f_{5/2}$ & $2f_{7/2}$ & $1h_{9/2}$ &
   $1i_{13/2}$ & & $3s_{1/2}$ & $2d_{3/2}$ & $2d_{5/2}$ & $1g_{7/2}$ & $1h_{11/2}$\\
\hline
$^{194}$Au & $\epsilon_{j_\nu}$ & 0.000 & 0.913 & 1.013 & 3.804 & 4.238
			   & 1.502 & $\epsilon_{j_\pi}$ & 0.000 & 0.915 & 2.640 &
					       5.165 &0.840 \\
           & $v^2_{j_\nu}$ & 0.254 & 0.521 & 0.555 & 0.945 & 0.950 &
			       0.699 & $v^2_{j_\pi}$ & 0.617 & 0.871 & 0.969 &
					       0.989 & 0.864 \\
\hline
$^{196}$Au & $\epsilon_{j_\nu}$ & 0.000 &0.929 & 1.036 & 3.831 & 4.321
   & 1.535 & $\epsilon_{j_\pi}$ & 0.000 &0.898 & 2.608 & 5.159 & 0.838 \\
           & $v^2_{j_\nu}$ & 0.296 & 0.595 & 0.632 & 0.956 & 0.962 &
   0.767 & $v^2_{j_\pi}$ & 0.618 & 0.869 & 0.968 & 0.989 & 0.864 \\
\hline
$^{198}$Au & $\epsilon_{j_\nu}$ & 0.000 &0.949 & 1.059 & 3.861 & 4.405
			   & 1.570 & $\epsilon_{j_\pi}$ & 0.000 &0.877 & 2.575 &
					       5.145 &0.827 \\
           & $v^2_{j_\nu}$ & 0.346 & 0.675 & 0.714 & 0.967 & 0.972 &
			       0.831 & $v^2_{j_\pi}$ & 0.621 & 0.868 & 0.968 &
					       0.989 & 0.865 \\
\hline\hline
  \end{tabular}
 \end{center}
\end{table*}

Once all the parameters of the IBFFM-2 Hamiltonian are obtained, it
is diagonalised numerically in the basis $|L_\nu L_\pi(L);j_\nu
j_\pi(J):I\rangle$, using the computer program TWBOS \cite{yoshida_priv_comm}. 
$L_\nu$ ($L_{\pi}$) and $L$ are the angular momentum for neutron (proton)
bosons and the total angular momentum for the even-even boson core, 
respectively. Finally, $I$ stands for the total angular momentum of the coupled system.

\subsection{Transition operators}

Using the eigenstates of the IBFFM-2 Hamiltonian, we can determine
the electric quadrupole (E2) and magnetic dipole (M1) properties of the odd-odd nuclei. 
In the present framework, the E2 operator $\hat T^{(E2)}$ takes the following form: 
\begin{align}
 \label{eq:e2}
\hat T^{(E2)}&= e_\nu^B\hat Q_\nu + e_\pi^B\hat Q_\pi
-\frac{1}{\sqrt{5}}\sum_{\rho=\nu,\pi}\sum_{j_{\rho}j'_{\rho}} \nonumber \\
&\times(u_{j_{\rho}}u_{j'_{\rho}}-v_{j_{\rho}}v_{j'_{\rho}})\langle
j'_{\rho}||e^F_{\rho}r^2Y^{(2)}||j_{\rho}\rangle(a_{j_{\rho}}^\dagger\times\tilde a_{j'_{\rho}})^{(2)},
\nonumber \\
\end{align}
where $e^B_\rho$ and $e^F_{\rho}$
stand for the effective charges for the boson and fermion systems,
respectively. 
The fixed values $e^B_\nu=e^B_\pi=0.15$ $e$b, which are taken from
Ref.~\cite{nomura2013hg}, and $e^F_\nu=0.5$ $e$b and $e^F_\pi=1.5$ $e$b
are used.

The M1 transition operator $\hat T^{(M1)}$ reads: 
\begin{align}
 \label{eq:m1}
\hat T^{(M1)}&=\sqrt{\frac{3}{4\pi}}
\Big\{
g_\nu^B\hat L^B_\nu + g_\pi^B\hat
L^B_\pi
-\frac{1}{\sqrt{3}}\sum_{\rho=\nu,\pi}\sum_{jj'} \nonumber \\
&\times (u_{j_{\rho}}u_{j'_{\rho}}+v_{j_{\rho}}v_{j'_{\rho}})\langle
j'_{\rho}||g_l^\rho{\bf l}+g_s^\rho{\bf s}||j_{\rho}\rangle(a_{j_{\rho}}^\dagger\times\tilde
a_{j'_{\rho}})^{(1)}
\Big\}. \nonumber \\
\end{align}
In this expression, $g_\nu^B$ and $g_\pi^B$ are the $g$-factors for the neutron and
proton bosons, respectively. The fixed values $g_\nu^B=0\,\mu_N$ and
$g_\pi^B=1.0\,\mu_N$  \cite{yoshida1985,IBM} are used in this work. 
For the neutron (proton) $g$-factors, the usual Schmidt values 
$g_l^\nu=0\,\mu_N$ and $g_s^\nu=-3.82\,\mu_N$
($g_l^\pi=1.0\,\mu_N$ and $g_s^\pi=5.58\,\mu_N$) are used. 
The $g_s$ value for both the proton and neutron are quenched by 30
\%. 

We note that the forms of the operators $\hat T^{(E2)}$
(Eq.~(\ref{eq:e2})) and $\hat T^{(M1)}$ (Eq.~(\ref{eq:m1})) have been
used in previous IBFFM-2 calculations
\cite{brant1984,brant2006,yoshida2013}. 

As we show later, we have computed the $B(E2)$ and $B(M1)$ transition rates, the
spectroscopic quadrupole moment $Q(I)$, and the magnetic moment $\mu(I)$
for the odd-odd nuclei $^{194,196,198}$Au, using the computer code
TWBTRN \cite{yoshida_priv_comm}.

\section{Results and discussion \label{sec:results}}


\begin{figure}[htb!]
\begin{center}
\includegraphics[width=\linewidth]{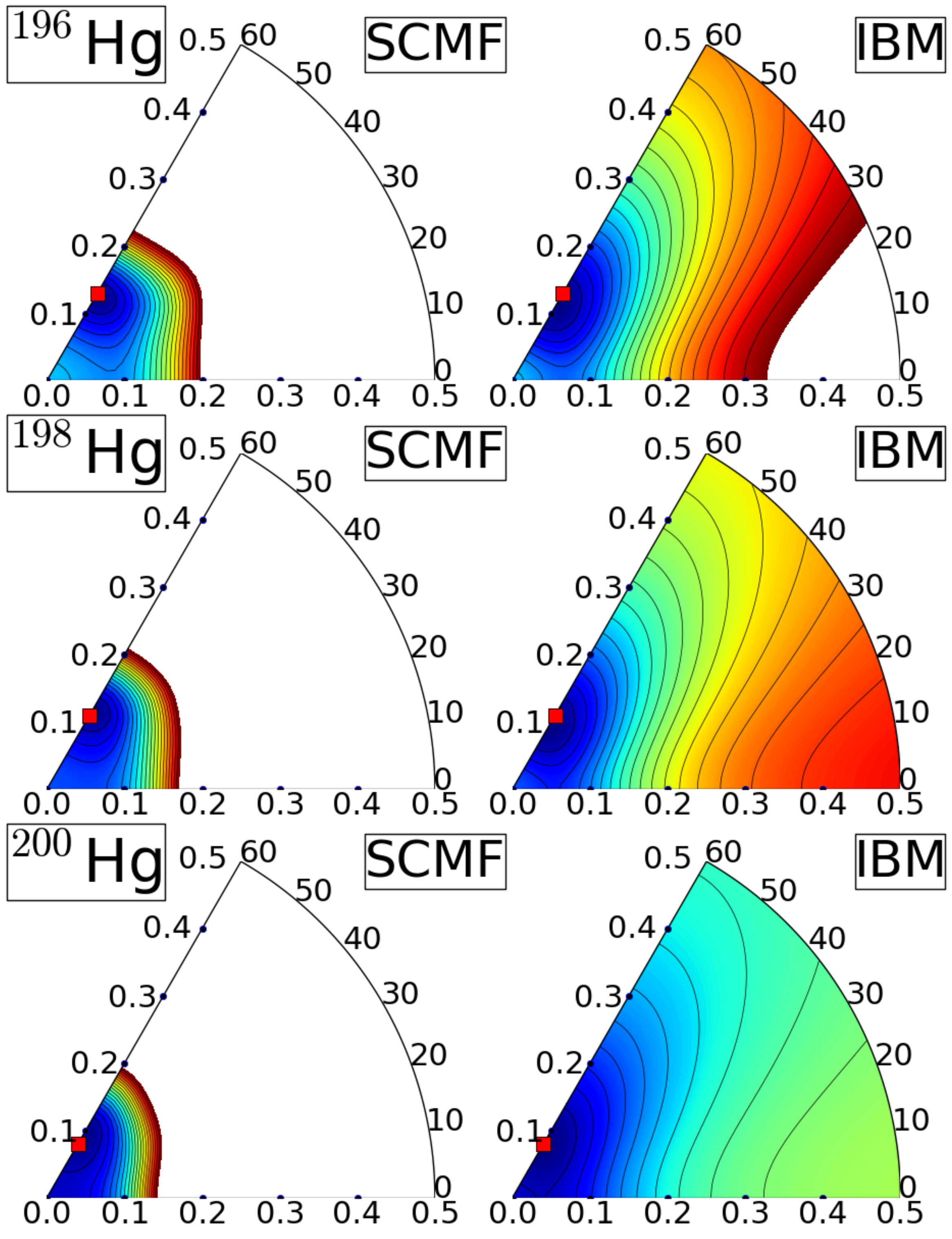}
\caption{(Color online) The Gogny-D1M HFB and mapped IBM-2 energy surfaces in the
 $(\beta,\gamma)$-deformation space for the 
 $^{196-200}$Hg  nuclei are plotted  up to 5 MeV from the
 global minimum. The energy difference between the neighbouring
 contours is 250 keV. The global minimum is indicated by a filled square.
 } 
\label{fig:pes}
\end{center}
\end{figure}

\subsection{Even-even Hg isotopes}


\begin{figure*}[htb!]
\begin{center}
\includegraphics[width=0.31\linewidth]{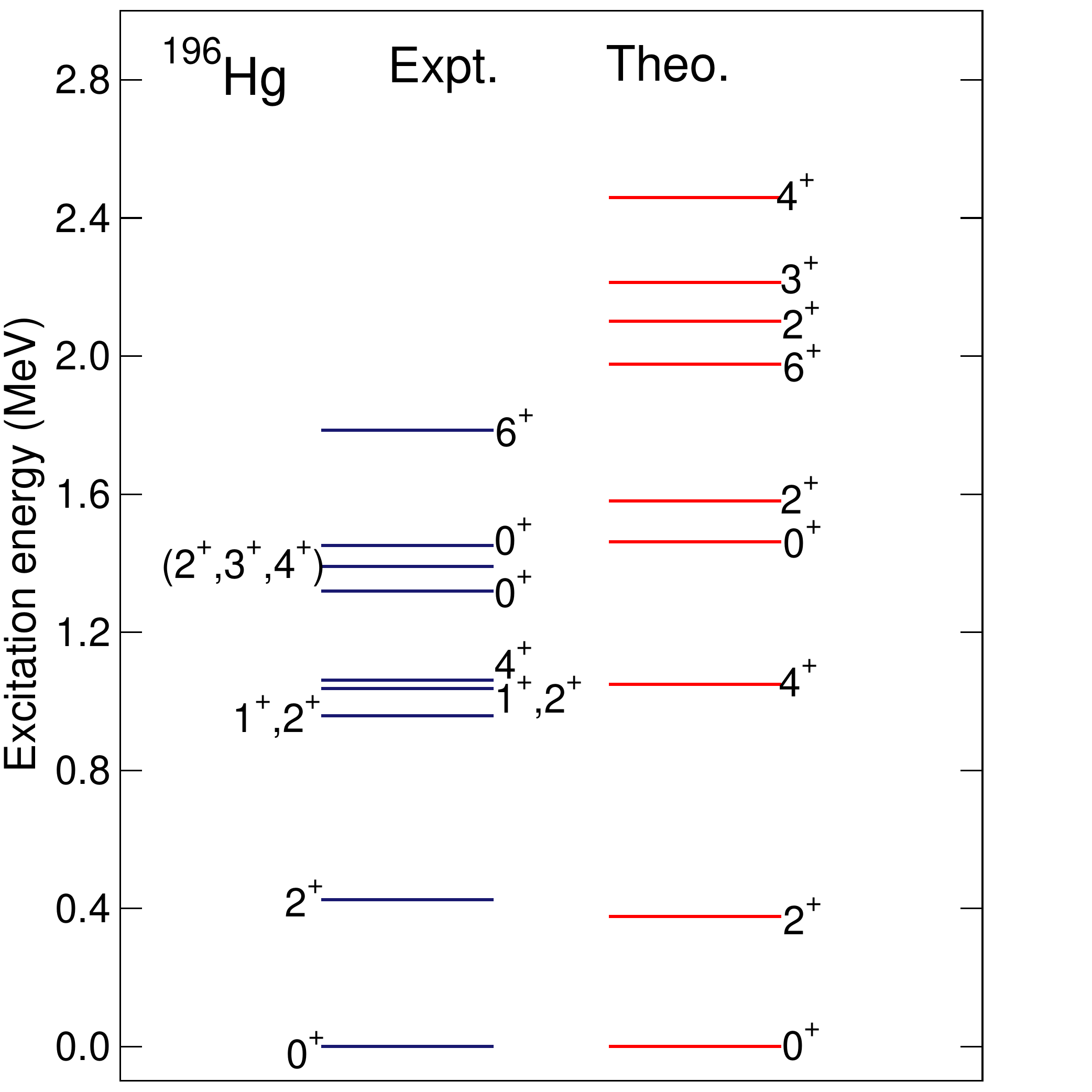} 
\includegraphics[width=0.31\linewidth]{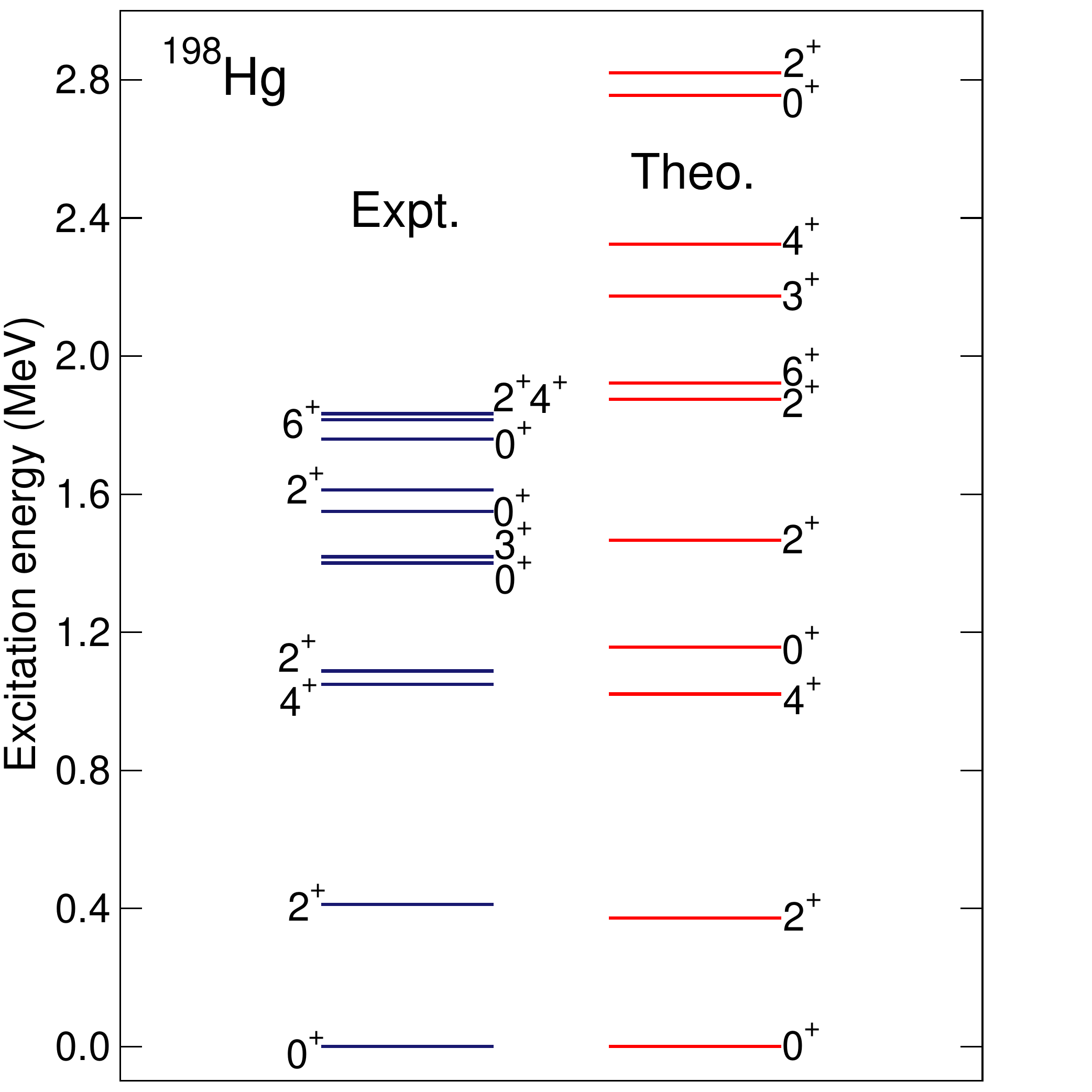} 
\includegraphics[width=0.31\linewidth]{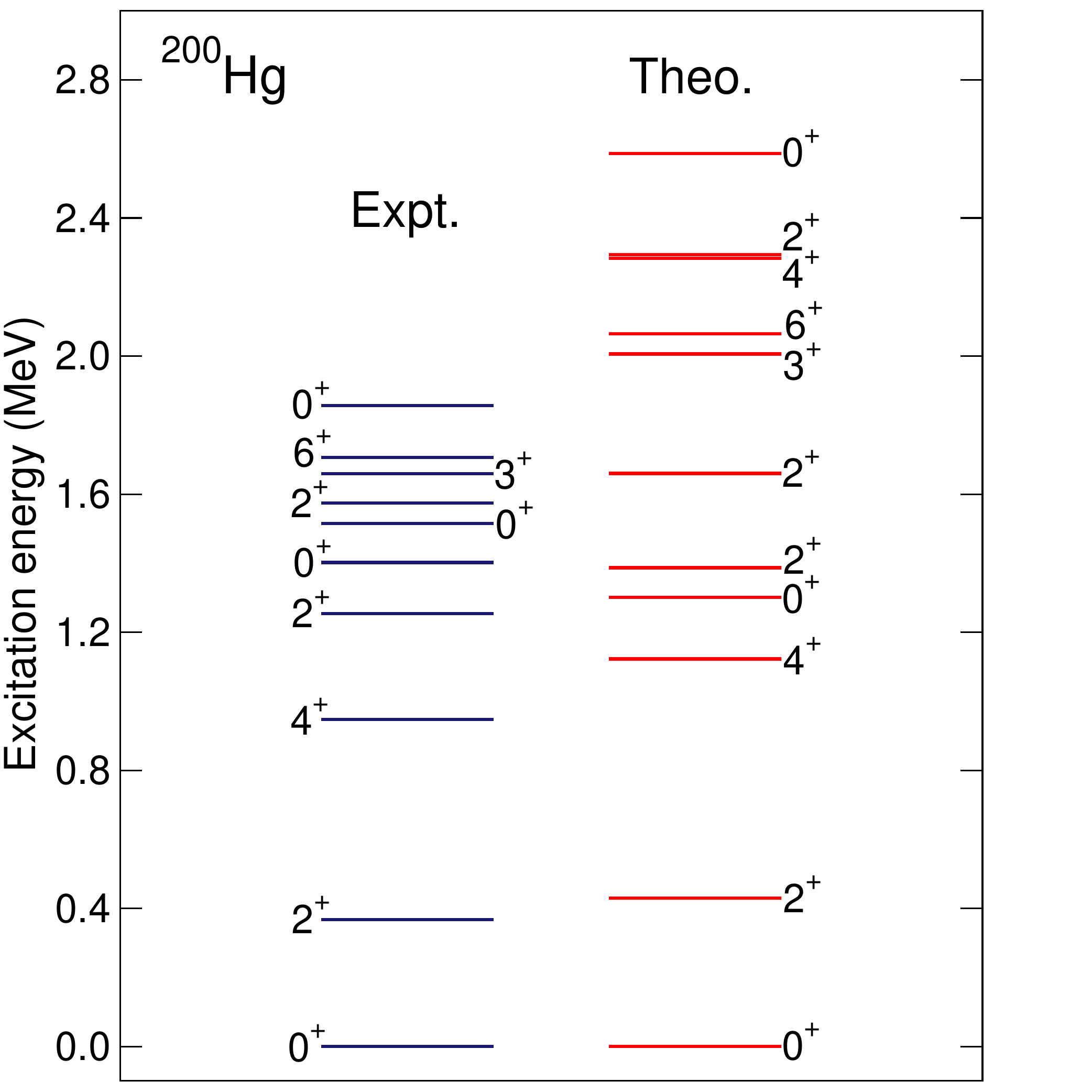} 
\caption{(Color online) Theoretical and experimental low-energy excitation spectra for
 the even-even nuclei $^{196,198,200}$Hg. The experimental values are taken from NNDC compilation \cite{data}.}
\label{fig:even-hg} 
\end{center}
\end{figure*}

In this section, we briefly discuss relevant results for the even-even
nuclei $^{196,198,200}$Hg, which were already presented in Ref.~\cite{nomura2013hg}. 
We plot in Fig.~\ref{fig:pes} the Gogny-D1M and mapped IBM-2 energy
surfaces for the $^{196,198,200}$Hg nuclei. 
In Fig.~\ref{fig:pes}, the Gogny-D1M energy surface for the $^{196}$Hg
nucleus exhibits a single oblate minimum located at $\beta\approx 0.13$. 
The oblate minimum becomes less pronounced in $^{198}$Hg and, finally,
the $^{200}$Hg nucleus exhibits a near spherical shape with a very shallow
oblate minimum at $\beta\approx 0.08$. 
The mapped IBM-2 energy surfaces on the right-hand side in
Fig.~\ref{fig:pes} reproduce the basic features of the original Gogny-D1M
ones around the global minimum, but look rather flat in the region away from the minimum. This is due to
the restricted boson model space \cite{nomura2008}, which only comprises a finite number of
bosons.

The calculated and experimental \cite{data} low-lying spectra are shown
in Fig.~\ref{fig:even-hg}. The yrast levels for all the
considered even-even nuclei are described reasonably well. 
However, the theoretical energy levels, in particular for $^{196,198}$Hg,
look more stretched than the experimental ones. 
For instance, our calculation is not able to account for the
excitation energy of the low-lying $0^+_3$ level of $^{196,198}$Hg. 
This discrepancy could be remedied by including in the IBM-2 model space
the intruder configurations that are associated with coexisting mean-field
minima. 
These configurations are, however, not considered for the nuclei studied 
here \cite{nomura2013hg}, since their corresponding Gogny-D1M energy surfaces only exhibit a
single mean-field minimum (see, Fig.~\ref{fig:pes}).

\subsection{Odd-mass Hg and Au isotopes}

\begin{figure}[htb!]
\begin{center}
\includegraphics[width=\linewidth]{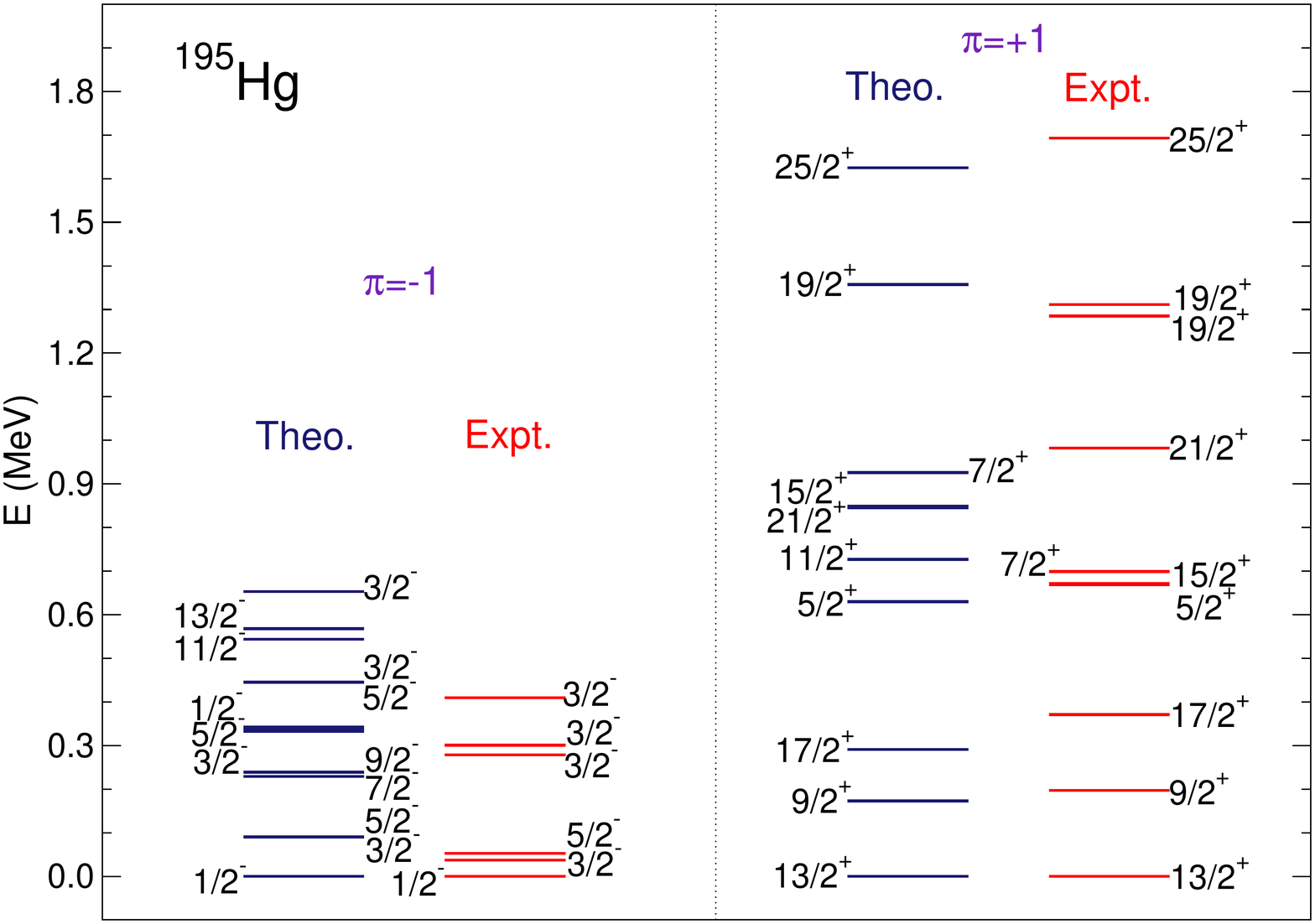}
\includegraphics[width=\linewidth]{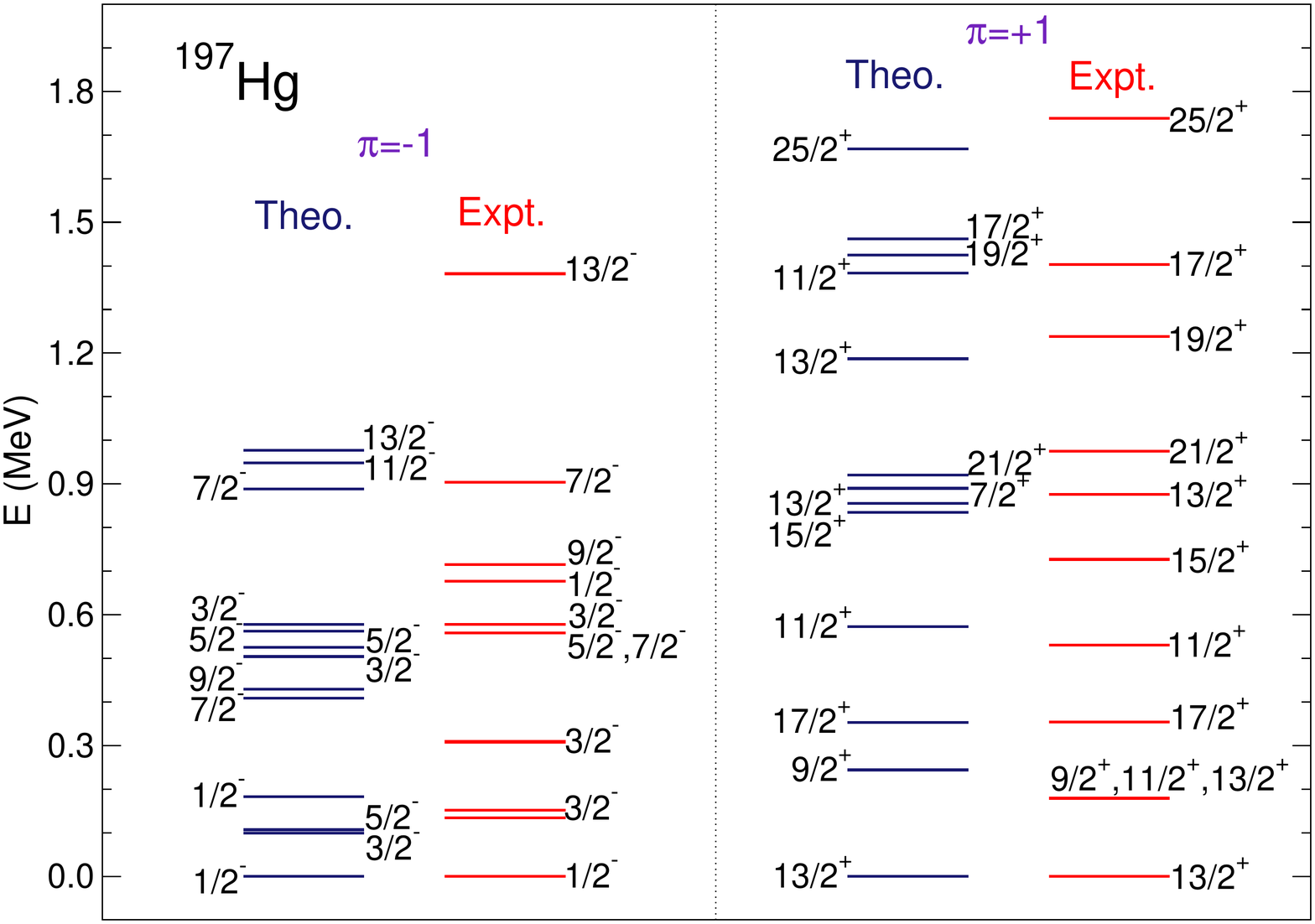} 
\includegraphics[width=\linewidth]{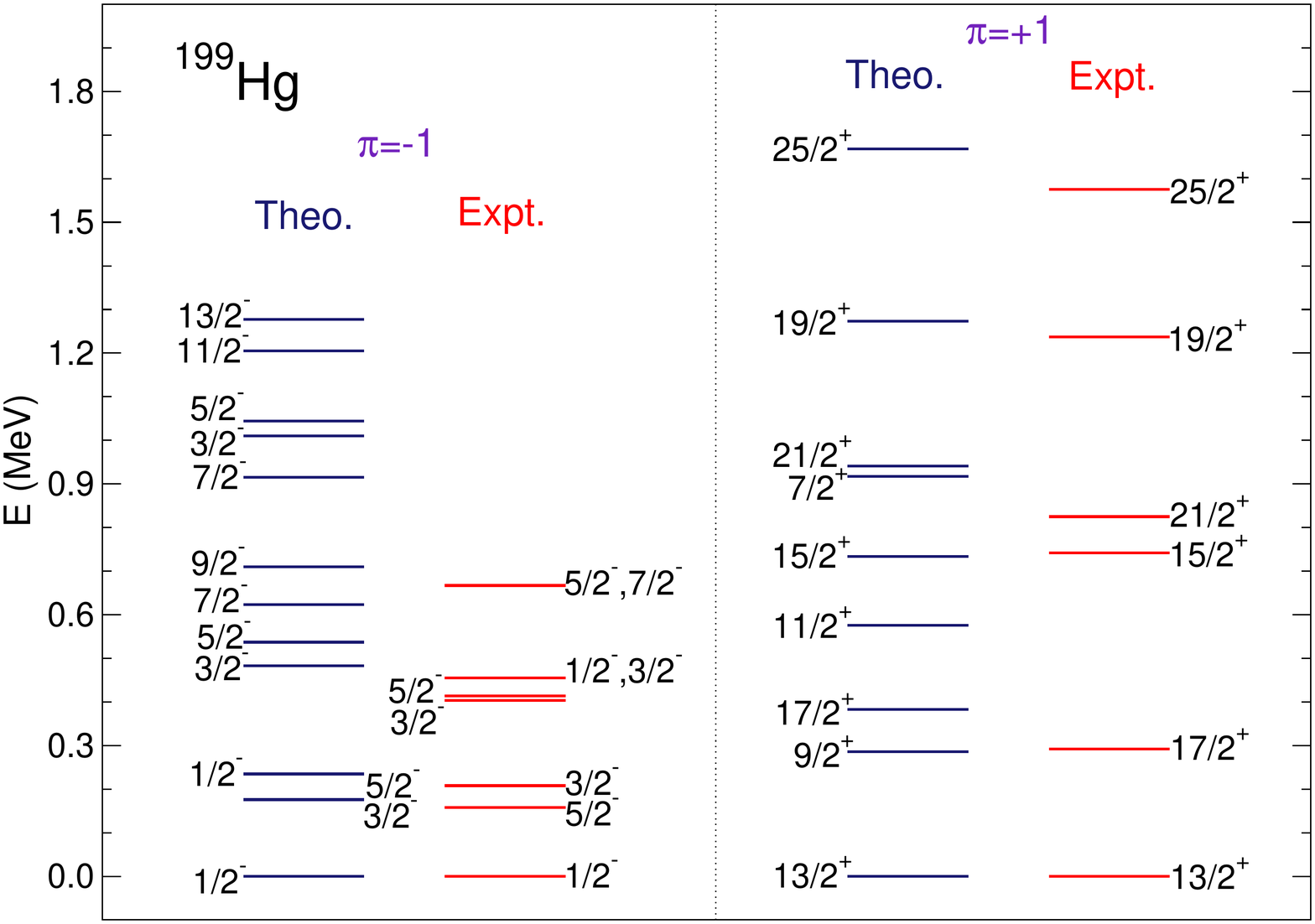} 
\caption{Same as Fig.~\ref{fig:even-hg}, but for the odd-N nuclei $^{195,197,199}$Hg.}
\label{fig:odd-hg} 
\end{center}
\end{figure}

\begin{figure}[htb!]
\begin{center}
\includegraphics[width=\linewidth]{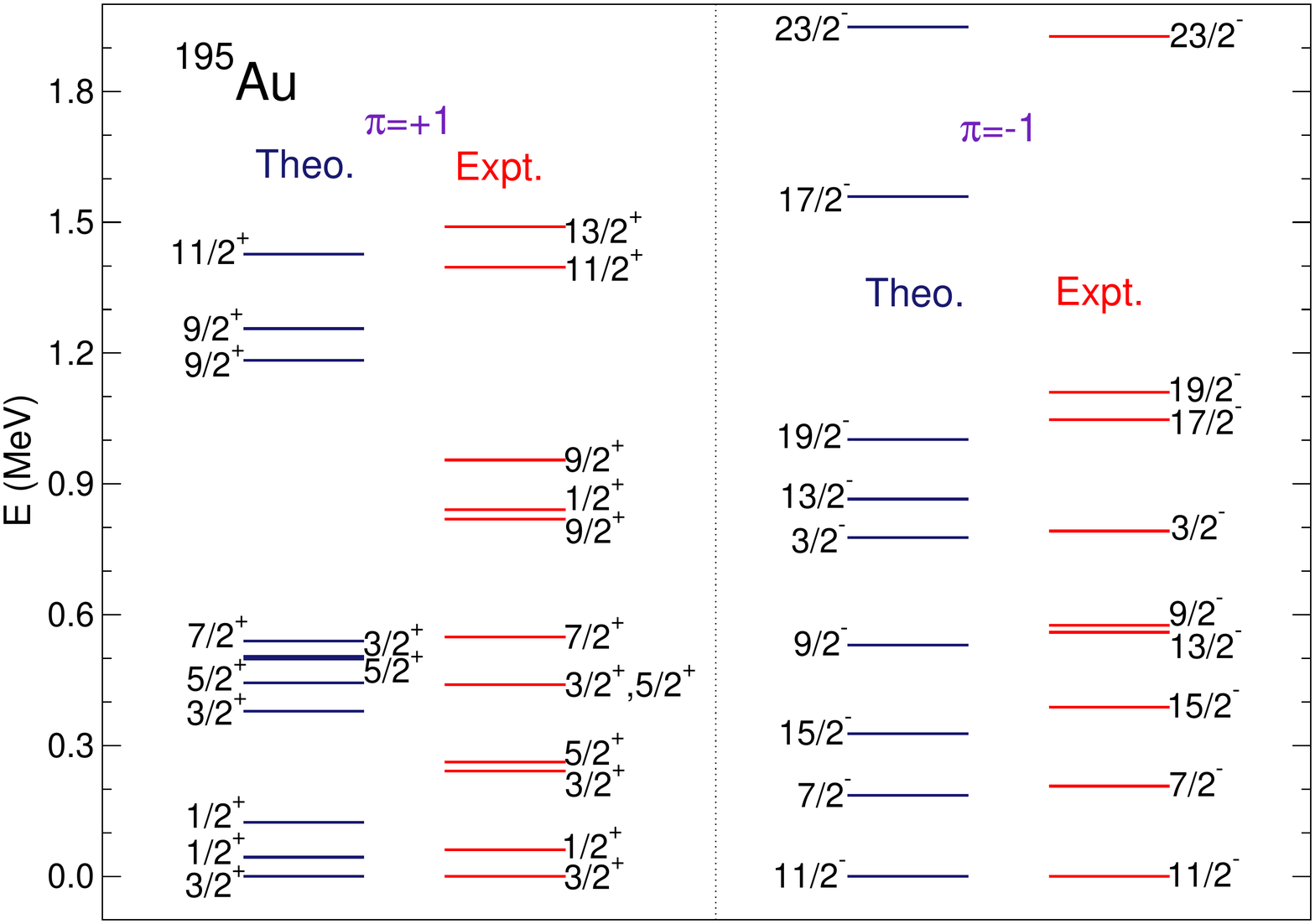}
\includegraphics[width=\linewidth]{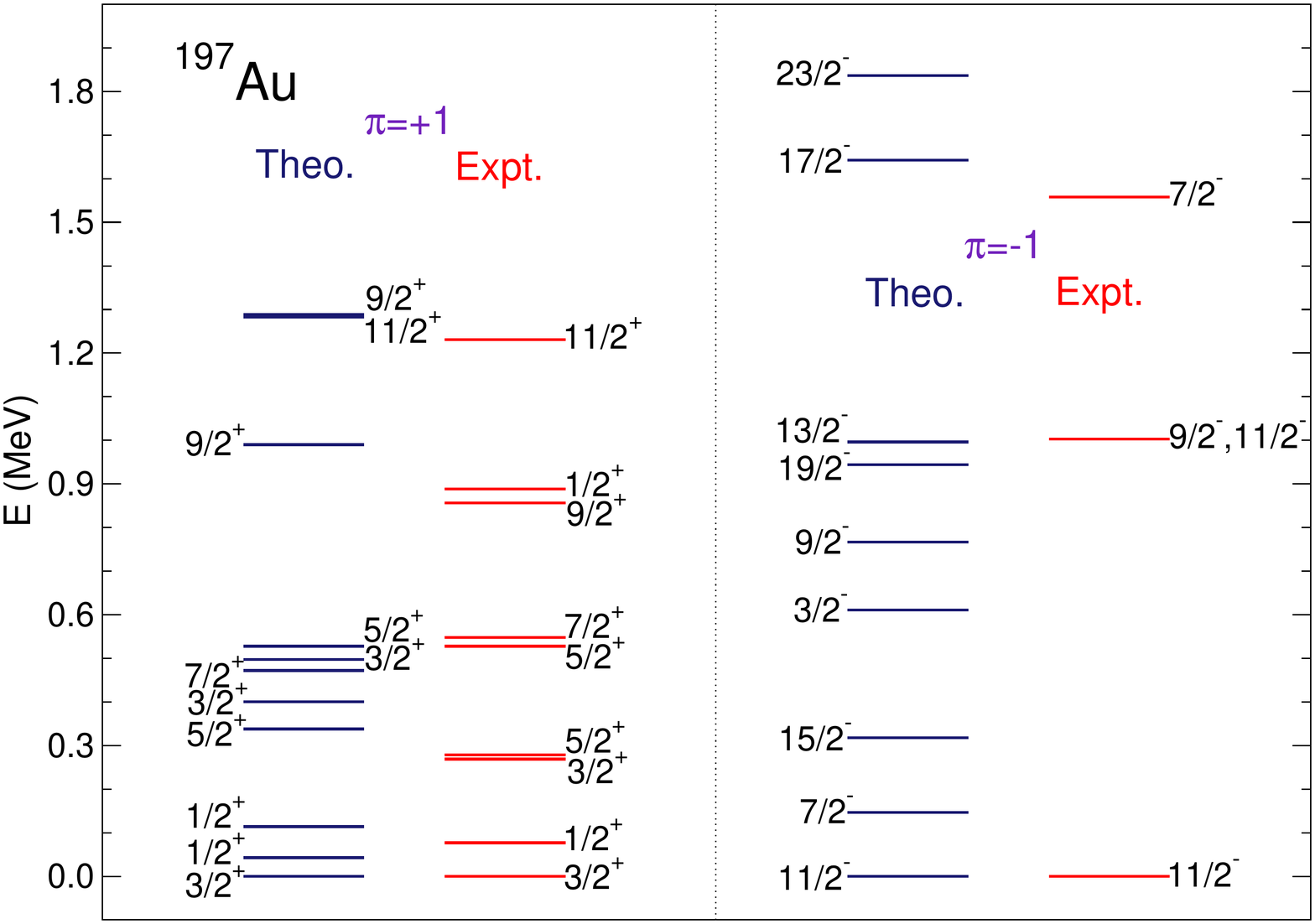} 
\includegraphics[width=\linewidth]{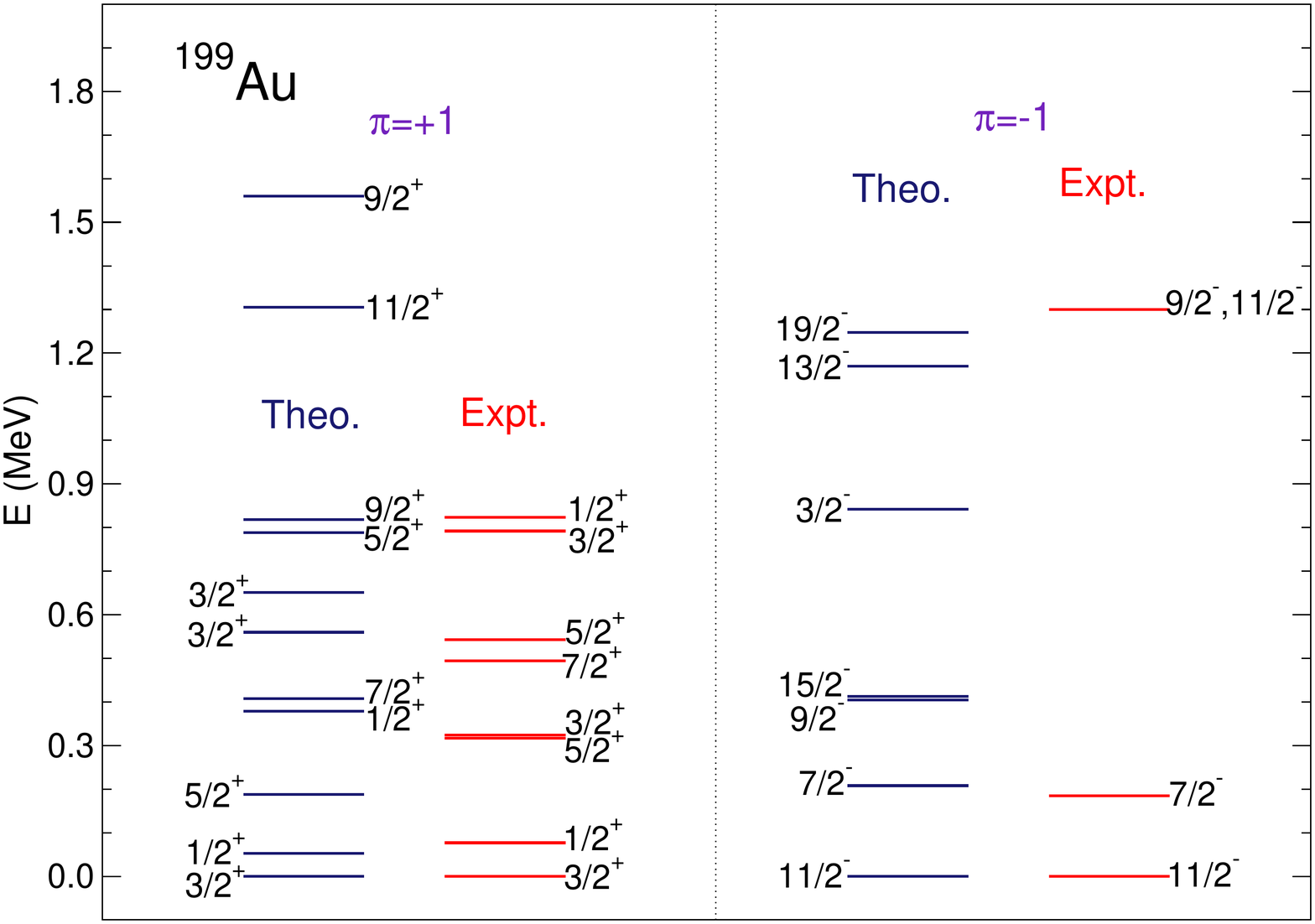} 
\caption{Same as Fig.~\ref{fig:even-hg}, but for the odd-Z nuclei $^{195,197,199}$Au.}
\label{fig:odd-au} 
\end{center}
\end{figure}

Next, we discuss the spectroscopic properties of the odd-mass nuclei,
obtained within the neutron-proton IBFM (IBFM-2). 
For the diagonalization of the IBFM-2 Hamiltonian, the computer code PBOS is used. 
The theoretical and experimental low-energy spectra for the odd-N nuclei
$^{195,197,199}$Hg are compared in Fig.~\ref{fig:odd-hg}. 
Especially for the positive-parity states, which are based on the unique-parity
$\nu i_{13/2}$ configuration, the present calculation
provides an excellent description of the experimental spectra for the
considered odd-N nuclei, although only three parameters are involved (see, Table~\ref{tab:bf-oddz}). 
The calculation reproduces nicely the ground-state band built on the
${13/2}^+_1$ state, which follows the $\Delta I=2$ systematic of the
weak-coupling limit. 
For the $^{195,197}$Hg nuclei, the calculation suggests that the
negative-parity yrast states near the ground state are based mainly on the odd neutron in the
$3p_{1/2}$ single-particle orbital coupled to the IBM-2 core. 
In the case of the nucleus $^{199}$Hg, however, in most of the yrast
states in the vicinity of the ground state three configurations
$3p_{1/2}$, $3p_{3/2}$, and $2f_{5/2}$ are more strongly mixed than in $^{195,197}$Hg. 
Such a change in the structure of the low-lying state from $^{195,197}$Hg
to $^{199}$Hg, reflects the evolution of shapes in the
corresponding even-even systems from $^{198}$Hg (weakly oblate
deformed) to $^{200}$Hg (nearly spherical).

In Fig.~\ref{fig:odd-au} we show similar plots for the odd-Z isotopes
$^{195,197,199}$Au. 
In general, our calculation is in a very good agreement with the
experimental data. 
Our calculation suggests that  the 
IBFM-2 wave functions of the lowest positive-parity states for
the considered odd-Z Au nuclei are composed, with a probability of more than 80 \%, of the 
$3s_{1/2}$ and $2d_{3/2}$ single-particle configurations, which are
substantially mixed with each other. 
On the other hand, the $2d_{5/2}$ and $1g_{7/2}$ configurations turn
out to play minor roles in describing the lowest-lying states.

We confirm that both the E2 and M1 properties of the
considered odd-mass nuclei are reasonably described with the present approach.

\subsection{Odd-odd Au isotopes}


\begin{figure}[htb!]
\begin{center}
\includegraphics[width=\linewidth]{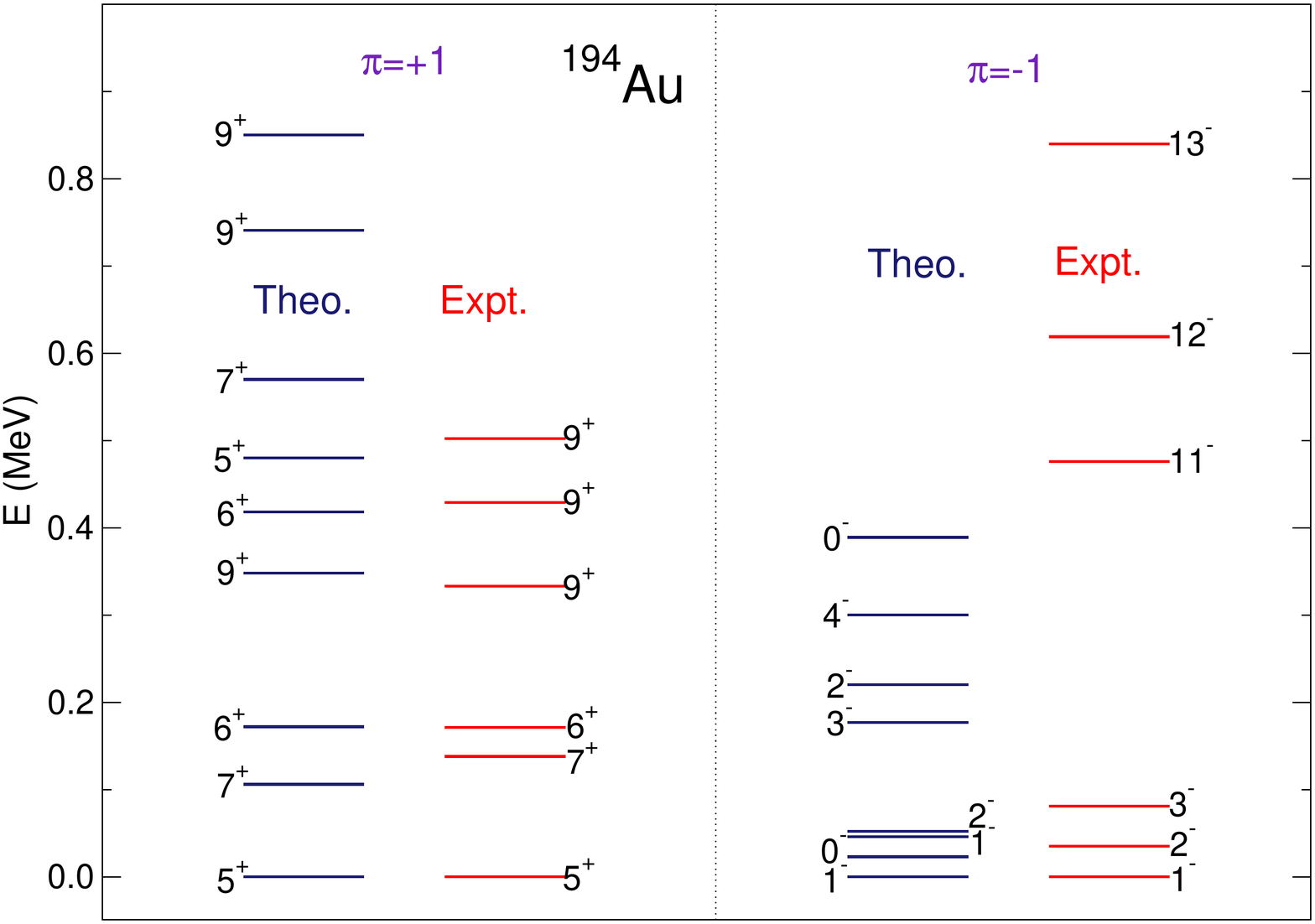}
\includegraphics[width=\linewidth]{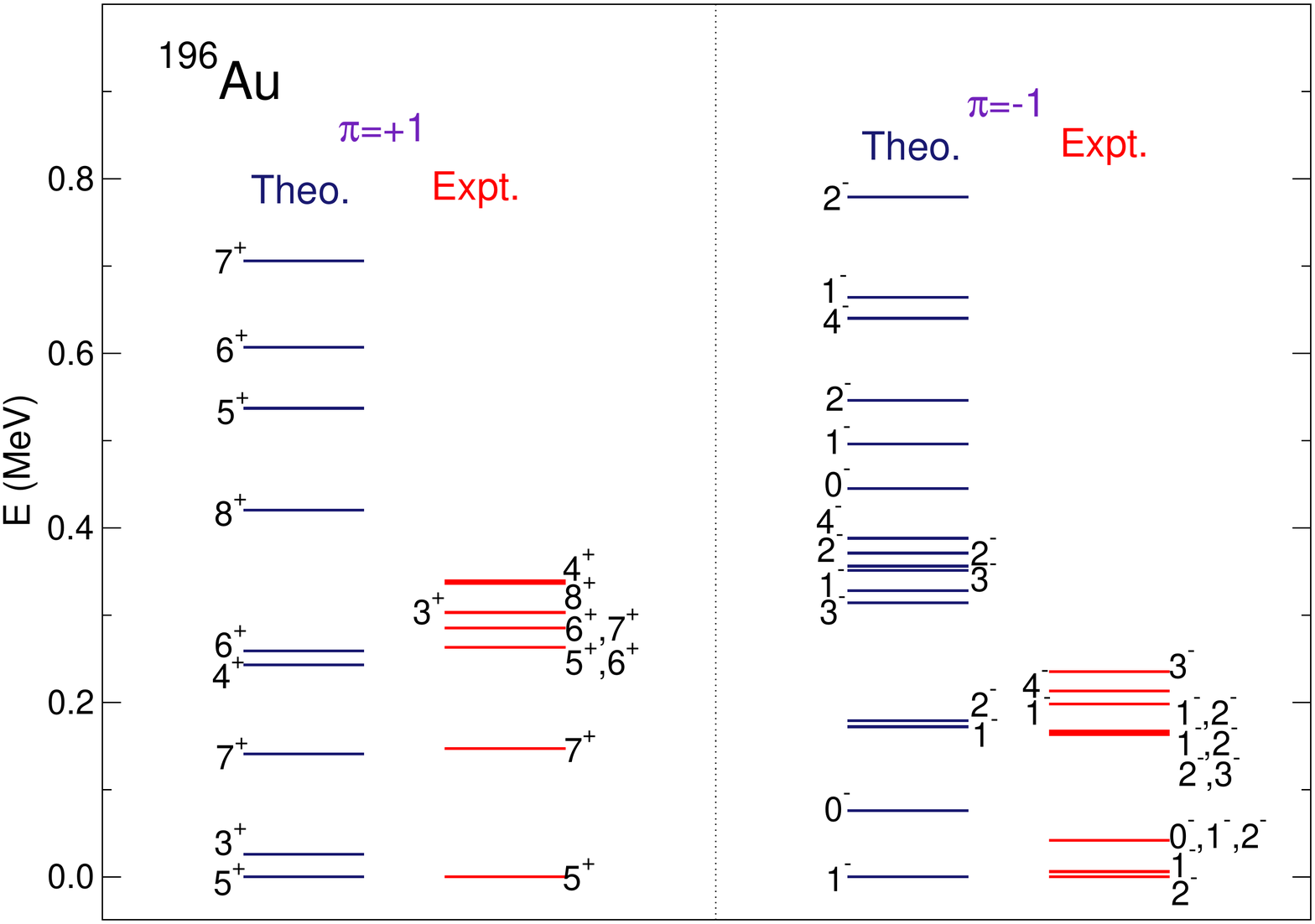} 
\includegraphics[width=\linewidth]{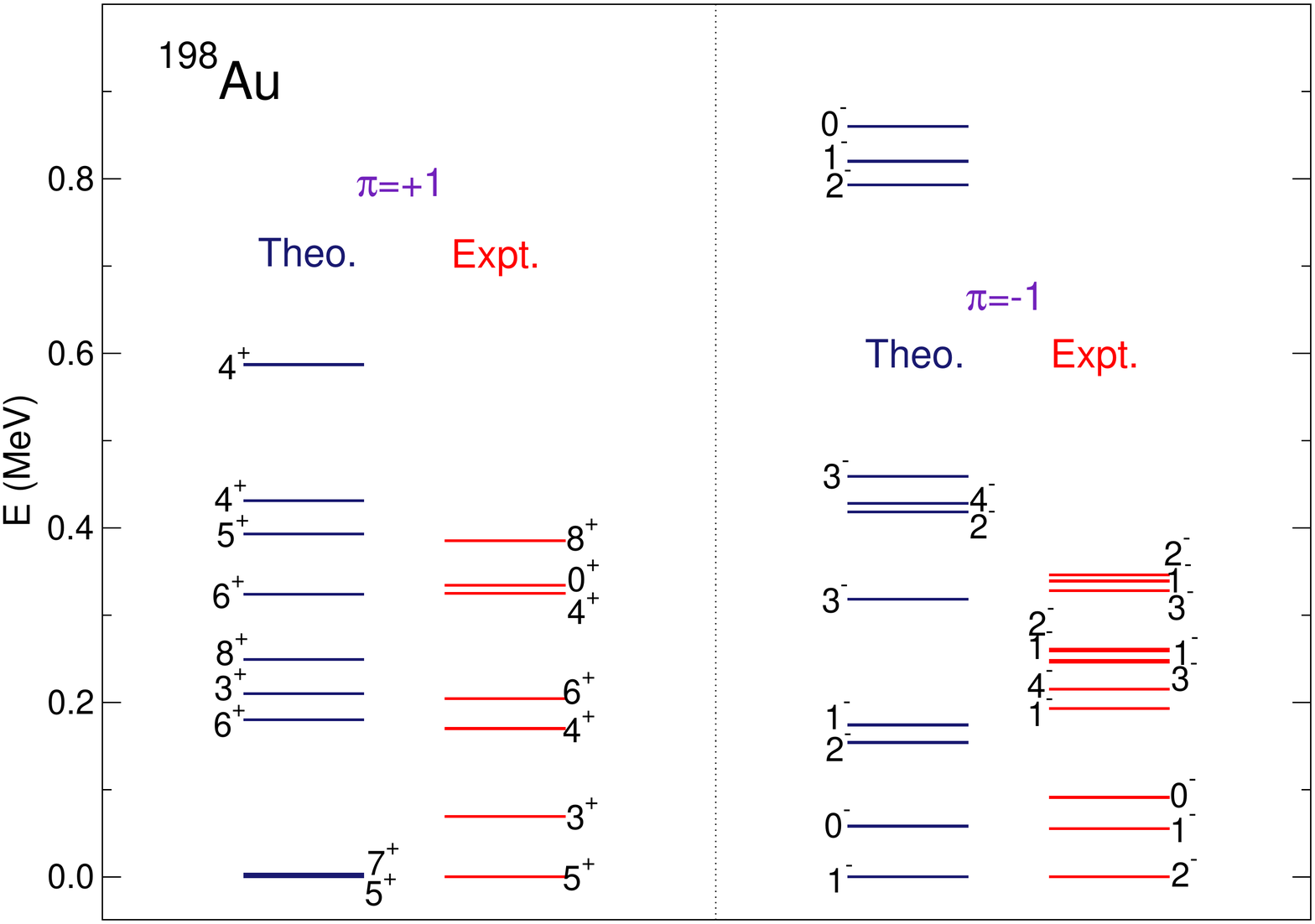}
\caption{Same as Fig.~\ref{fig:even-hg}, but for the odd-odd nuclei
 $^{194,196,198}$Au. The experimental values are taken from
 Refs.~\cite{data,gao2012} (for $^{194}$Au), \cite{data,petkov2007}
 (for $^{196}$Au), and \cite{data} (for $^{198}$Au)}
\label{fig:dodd-au} 
\end{center}
\end{figure}

Let us now focus on the discussion of the results for the odd-odd nuclei. 
The low-lying spectra computed with the IBFFM-2 for the odd-odd
$^{194,196,198}$Au nuclei are depicted in Fig.~\ref{fig:dodd-au}, and
compared with the experimental data \cite{data}.

\subsubsection{$^{194}$Au}

Firstly, we observe that the present IBFFM-2 result for the $^{194}$Au nucleus
is in a very good agreement with the experimental spectra, especially
for the positive-parity states. 
The main component ($\approx$ 72 \%) in the wave functions of the lowest
three positive-parity states, i.e., ${5^+_1}$, $7^+_1$, and $6^+_1$ states, is the $(\nu p_{1/2}\otimes\pi
h_{11/2})$ neutron-proton pair coupled to the boson core. 
For the negative-parity, the energy levels near the $1^-$ ground state
are reasonably reproduced in the present calculation. The main
components of IBFFM-2 wave function of the $1^-$ state are the 
$(\nu p_{1/2}\otimes\pi s_{1/2})_{J=1^-}$ (17 \%), and 
$(\nu f_{5/2}\otimes\pi s_{1/2})_{J=3^-}$ neutron-proton pairs (13 \%). 
However, the calculation is not able to reproduce the experimental $11^-$, $12^-$, 
and $13^-$ levels, which are below 1 MeV excitation. The excitation energies for
these states are predicted to be much larger ($>$ 3 MeV).
Empirically, these higher-spin negative parity states are mainly made of
the pair composed of the unique-parity orbitals, i.e., $(\nu
i_{13/2}\otimes\pi h_{11/2})$ \cite{data}.
The corresponding IBFFM-2 wave functions obtained in the present work
are, however, made of the coupling between the odd neutron and 
proton in the normal-parity orbitals: for instance, the main components
of the predicted $11^-_1$ states are the 
$(\nu p_{1/2}\otimes\pi s_{1/2})_{J=1^-}$ (35 \%), 
$(\nu f_{5/2}\otimes\pi s_{1/2})_{J=3^-}$ (19 \%), and $(\nu
p_{1/2}\otimes\pi d_{5/2})_{J=3^-}$ (13 \%) neutron-protons pairs.

\begin{table}[!htb]
\begin{center}
\caption{\label{tab:194au} Calculated and experimental $B(E2)$ and
 $B(M1)$ transition rates (in Weisskopf units), and quadrupole $Q(I)$
 (in $e$b units) and magnetic $\mu(I)$ (in $\mu_N$ units) moments for
 the odd-odd nucleus $^{194}$Au. The experimental values are taken
 from Ref.~\cite{data}.}
 \begin{tabular}{ccc}
\hline\hline
  & Theory & Experiment \\
\hline
$B(E2; 6^+_1\rightarrow 5^+_1)$ & 3.0 & 5(3) \\
$B(E2; 7^+_1\rightarrow 5^+_1)$ & 61 & 27(2) \\
$B(E2; 8^+_1\rightarrow 6^+_1)$ & 61 & 22(4) \\
$B(E2; 8^+_1\rightarrow 7^+_1)$ & 1.5 & 1.9(5) \\
$B(M1; 6^+_1\rightarrow 5^+_1)$ & 0.050 & 0.0010(4) \\
$B(M1; 6^+_1\rightarrow 7^+_1)$ & 0.020 & 0.0034(14) \\
$B(M1; 8^+_1\rightarrow 7^+_1)$ & 0.021 & 5$\times 10^{-5}$(2) \\
$Q(1^-_1)$ & +0.225 & -0.240(9) \\
$\mu(1^-_1)$ & +1.790 & +0.0763(13) \\
\hline\hline
 \end{tabular}
\end{center} 
\end{table}

The experimental information about the electromagnetic properties is
rather scarce for $^{194}$Au. Nevertheless, we show in
Table~\ref{tab:194au} the calculated $B(E2)$ and $B(M1)$ transition
rates, and quadrupole $Q(I)$ and magnetic $\mu(I)$ moments  in
comparison with the available data. 
The predicted $B(E2)$ values seem to be qualitatively in a good
agreement with the data. 
The calculated $B(M1)$ values are, however, too large as compared
to the experimental values. 
The sign of the predicted $Q(1^-)$  moment is opposite
to that given by the experiment. 


\subsubsection{$^{196}$Au}

From the comparison of energy levels of the odd-odd nucleus $^{196}$Au, shown in
Fig.~\ref{fig:dodd-au}, one concludes that the calculation is able to reproduce
both the experimental positive- and negative-parity levels reasonably well. 
The calculated low-lying positive-parity states for $^{196}$Au are similar in
structure to those for $^{194}$Au: 67 \% and 66 \% of the predicted
$5^+_1$ and $7^+_1$ states are dominated by the 
$(\nu p_{1/2}\otimes h_{11/2})_{J=5^+}$ pair component, respectively. 
The spin of the calculated lowest negative parity state is
$I=1^-$. This is at variance with the experiment, although the
experimental $1^-_1$ level is only 6 keV above the $2^-_1$ ground
state. 
Furthermore, the present calculation considerably overestimates the $2^-_1$ energy level.
The non-negligible components ($>$ 10 \%) of the corresponding IBFFM-2 wave functions for the
$2^-_1$ and $1^-_1$ states are
the following: 
$(\nu p_{1/2}\otimes\pi d_{3/2})_{J=2^-}$ (24 \%),  
$(\nu p_{1/2}\otimes\pi s_{1/2})_{J=1^-}$ (11 \%), and 
$(\nu f_{5/2}\otimes\pi s_{1/2})_{J=2^-}$ (11 \%) for the $2^-_1$ state, and 
$(\nu p_{1/2}\otimes\pi s_{1/2})_{J=1^-}$ (38 \%) and 
$(\nu p_{1/2}\otimes\pi d_{3/2})_{J=1^-}$ (10 \%) for the $1^-_1$ state. 

\begin{table}[!htb]
\begin{center}
\caption{\label{tab:196au} Same as Table~\ref{tab:194au}, but for the
 nucleus $^{196}$Au. The experimental values are taken from Refs.~\cite{data,petkov2007}}
 \begin{tabular}{ccc}
\hline\hline
  & Theory & Experiment \\
\hline
$B(E2; 6^+_1\rightarrow 5^+_1)$ & 2.1 & $>$0.064 \\
$B(E2; 6^+_2\rightarrow 5^+_1)$ & 7.7 & $>$0.0068 \\
$B(E2; 7^+_1\rightarrow 5^+_1)$ & 50 & 51(6) \\
$B(E2; 7^+_2\rightarrow 5^+_1)$ & 0.24 & $>$0.064 \\
$B(E2; 8^+_1\rightarrow 6^+_1)$ & 51 & 0.77(39) \\
$B(E2; 8^+_1\rightarrow 6^+_2)$ & 0.28 & 20(7) \\
$B(E2; 8^+_1\rightarrow 7^+_1)$ & 1.0 & 0.76$\times 10^{-1}$(26) \\
$B(E2; 8^+_1\rightarrow 7^+_2)$ & 0.018 & 0.77(39) \\
$B(E2; 3^-_2\rightarrow 1^-_1)$ & 11 & $>$6.5 \\
$B(E2; 4^-_1\rightarrow 2^-_1)$ & 21 & 9.7(2.4) \\
$B(E2; 4^-_1\rightarrow 2^-_2)$ & 1.8 & 13.2(13.2) \\
$B(E2; 4^-_1\rightarrow 3^-_1)$ & 0.34 & 13.2(13.2) \\
$B(M1; 6^+_1\rightarrow 7^+_1)$ & 0.029 & 3.5$\times 10^{-5}$ \\
$B(M1; 6^+_2\rightarrow 6^+_1)$ & 0.077 & $>$0.00016 \\
$B(M1; 6^+_2\rightarrow 7^+_1)$ & 0.14 & 3.5$\times 10^{-5}$ \\
$B(M1; 6^+_2\rightarrow 7^+_2)$ & 0.050 & $>$0.00016 \\
$B(M1; 7^+_2\rightarrow 7^+_1)$ & 0.0053 & 3.5$\times 10^{-5}$ \\
$B(M1; 8^+_1\rightarrow 7^+_1)$ & 0.022 & 0.49$\times 10^{-3}$(5) \\
$B(M1; 3^-_2\rightarrow 2^-_1)$ & 0.17  & $>$0.0045 \\
$Q(2^-_1)$ & +0.495 & 0.81(7) \\
$\mu(2^-_1)$ & +0.197 & +0.580(15) \\
\hline\hline
 \end{tabular}
\end{center} 
\end{table}

Table~\ref{tab:196au} exhibits the calculated and experimental
electromagnetic properties. Regarding the $B(E2)$ rates, our results are
in a reasonable agreement with the experiment. 
However, similarly to $^{196}$Au the calculated $B(M1)$ values are
generally much larger than the experimental values.
A number of experimental $B(E2)$ and $B(M1)$ transition rates from the $1^-$
state at the excitation energy of $E_\text{x}=298.5$ keV are available \cite{petkov2007}. 
However, there are also too many experimental $1^-$ states below 298.5 keV, and it 
is not clear which theoretical $1^-$ state corresponds to the
experimental one observed at $E_\text{x}=298.5$ keV. 
For this reason, we do not compare our results with the 
experimental $B(E2)$ and $B(M1)$ transitions rates from 
the $1^-(298.5\,\text{keV})$ state.

\subsubsection{$^{198}$Au}

As one sees from the comparison between the theoretical and experimental
low-energy spectra for the odd-odd nucleus $^{198}$Au in
Fig.~\ref{fig:dodd-au}, the description of the positive-parity states is
generally good. 
As in the case of $^{196}$Au, however, our calculation fails to
reproduce the spin of 
the lowest negative-parity state. 
The structure of the $2^-_1$ and $1^-_1$ wave functions for $^{198}$Au
turn out to be rather similar to those of $^{196}$Au, that is, 
$(\nu p_{1/2}\otimes\pi d_{3/2})_{J=2^-}$ (26 \%), 
$(\nu p_{1/2}\otimes\pi s_{1/2})_{J=1^-}$ (19 \%), and 
$(\nu p_{1/2}\otimes\pi s_{1/2})_{J=0^-}$ (12 \%) for the $2^-_1$ state, and 
$(\nu p_{1/2}\otimes\pi s_{1/2})_{J=1^-}$ (41 \%) and 
$(\nu p_{1/2}\otimes\pi d_{3/2})_{J=1^-}$ (13 \%) for the $1^-_1$ state. 
The previous IBFFM calculation of \cite{lopac1986} obtains an excellent
description of both the positive- and negative-parity levels. 
The IBFFM wave functions they obtained are predominantly described by the $(\nu p_{1/2}\otimes\pi
d_{3/2})_{J=2^-}$ component ($>$ 70 \%) for the $2^-_1$ state, and $(\nu
p_{1/2}\otimes\pi d_{3/2})_{J=1^-}$ (50 \%) for the $1^-_1$ state. 
The difference between our result and that of \cite{lopac1986} could be
accounted for by the different single-particle energies used in each study. 
In the present calculation, the $2d_{3/2}$ single-particle orbital is
about 0.9 MeV above the $3s_{1/2}$ (see, Table~\ref{tab:vsq-dodd}). 
On the other hand, in \cite{lopac1986} the $2d_{3/2}$ orbital 
is below the $3s_{1/2}$ one and, consequently, the $\pi d_{3/2}$
single-particle configuration plays a more dominant role in low-energy
region than in our calculation. 

\begin{table}[!htb]
\begin{center}
\caption{\label{tab:198au} Same as Table~\ref{tab:194au}, but for the
 nucleus $^{198}$Au.}
 \begin{tabular}{ccc}
\hline\hline
  & Theory & Experiment \\
\hline
$B(E2; 1^-_2\rightarrow 2^-_1)$ & 6.7 & 2.2(7) \\
$B(E2; 2^-_3\rightarrow 4^-_1)$ & 0.049 & $>$64 \\
$B(E2; 3^-_1\rightarrow 1^-_1)$ & 8.3 & $>$26 \\
$B(E2; 3^-_1\rightarrow 2^-_1)$ & 2.1 & $>$13 \\
$B(E2; 4^-_1\rightarrow 2^-_1)$ & 13 & 35(18) \\
$B(M1; 1^-_2\rightarrow 0^-_1)$ & 0.015 & 0.0032(10) \\
$B(M1; 1^-_2\rightarrow 1^-_1)$ & 0.0049 & 0.00024(8) \\
$B(M1; 1^-_3\rightarrow 0^-_1)$ & 0.00067 & 7.4$\times 10^{-5}$(24) \\
$B(M1; 1^-_3\rightarrow 1^-_1)$ & 0.00051 & 0.0017(5) \\
$B(M1; 1^-_4\rightarrow 0^-_1)$ & 8.2$\times 10^{-5}$ & $>$0.0084 \\
$B(M1; 1^-_4\rightarrow 2^-_1)$ & 0.0021 & $>$9.9$\times 10^{-6}$ \\
$B(M1; 1^-_5\rightarrow 1^-_1)$ & 0.022  & $>$0.00029 \\
$B(M1; 1^-_5\rightarrow 1^-_2)$ & 0.042 & $>$0.0042 \\
$B(M1; 1^-_6\rightarrow 0^-_1)$ & 0.0013 & $>$0.00025 \\
$B(M1; 1^-_6\rightarrow 1^-_4)$ & 0.10  & $>$0.015 \\
$B(M1; 1^-_6\rightarrow 2^-_1)$ & 0.0010 & $>$5.6$\times 10^{-5}$ \\
$B(M1; 2^-_2\rightarrow 1^-_1)$ & 0.064 & $>$0.00033 \\
$B(M1; 2^-_2\rightarrow 2^-_1)$ & 0.024 & $>$0.0037 \\
$B(M1; 2^-_3\rightarrow 1^-_2)$ & 4.5$\times 10^{-6}$ & $>$0.00065 \\
$B(M1; 2^-_3\rightarrow 1^-_3)$ & 0.024 & $>$0.0048 \\
$B(M1; 2^-_3\rightarrow 2^-_1)$ & 8.5$\times 10^{-5}$ & $>$0.00042 \\
$B(M1; 2^-_4\rightarrow 1^-_4)$ & 0.0015 & $>$0.015 \\
$B(M1; 3^-_1\rightarrow 2^-_1)$ & 0.42 & $>$0.0019 \\
$B(M1; 3^-_2\rightarrow 2^-_1)$ & 0.0042 & $>$0.0026 \\
$Q(2^-_1)$ & +0.373 & +0.64(2) \\
$\mu(5^+_1)$ & +4.398 & -1.11(2) \\
$\mu(2^-_1)$ & +0.334  & +0.5934(4) \\
\hline\hline
 \end{tabular}
\end{center} 
\end{table}

In Table~\ref{tab:198au} the calculated $B(E2)$ values for $^{198}$Au are, in
general, in good agreement with the experiment. 
We also present the calculated $B(M1)$, but for most of the
available data only a lower limit for this quantity is known. 
The calculated magnetic moment of the $5^+_1$ state, $\mu(5^+_1)$, has the opposite sign  and is a factor of 4
larger in magnitude than the experimental one. 
Similar results have been obtained for the $^{194,196}$Au
nuclei as we obtain $\mu(5^+_1)\approx 5$
$\mu_N$. As already mentioned, the $5^+_1$ states obtained in the
present calculation for the considered odd-odd Au nuclei are dominated by the $(\nu
p_{1/2}\otimes\pi h_{11/2})_{J=5^+}$ neutron-proton pair configuration,
and the predicted $\mu(5^+_1)$ moments are mostly accounted for by this
configuration, in particular, by the odd-proton part of the M1 matrix
element, which takes large positive value. 
On the other hand,  empirical studies for the low-lying level structure of
$^{194}$Au \cite{pakkanen1977,gao2012} assume the $5^+_1$ state and
the band built on it to be based mainly on the $(\nu i_{13/2}^{-1}\otimes\pi
d_{3/2}^{-1})_{J=5^+}$ configuration, leading to the correct sign of the
$\mu(5^+_1)$ moment.


\section{Summary and concluding remarks\label{sec:summary}}


In this work, we extend the recently developed method of
Ref.~\cite{nomura2016odd} for calculating the spectroscopy of odd-mass
nuclei to odd-odd systems. 
The $(\beta,\gamma)$-deformation energy surfaces of the even-even core
nuclei, and spherical single-particle energies and occupation
probabilities of the odd neutron and the odd proton are calculated
by the constrained HFB method based on the Gogny D1M EDF. 
These quantities are then used as microscopic input to build most
of the different terms of the IBFFM-2 Hamiltonian. 
The strength parameters for the boson-fermion interaction terms in the
IBFFM-2 Hamiltonian are taken from those of the neighbouring odd-mass
nuclei. 
Two coefficients in the residual interaction between odd neutron and
proton are the only new parameters, and are determined as to
reproduce the low-energy levels of each odd-odd nucleus. 
In this way, we are able to reduce significantly the number of
free parameters in the IBFFM-2 framework. 
It is shown that the method provides a reasonable description of
low-energy spectra and electromagnetic properties of the odd-odd nuclei
$^{194,196,198}$Au. 
Even though a few strength parameters in the boson-fermion and
fermion-fermion interactions are treated as free parameters, the method
developed in this paper, as well as in Ref.~\cite{nomura2016odd}, in which the
even-even IBM-core Hamiltonian is
determined fully microscopically and only one or two unpaired nucleon
degrees of freedom are added via the particle-boson coupling, allows for
a simultaneous description of a large number of even-even, odd-mass, and
odd-odd medium-mass and heavy nuclei.

\acknowledgments
This work was supported in part by the QuantiXLie Centre of Excellence, a project
co-financed by the Croatian Government and European Union through the
European Regional Development Fund - the Competitiveness and Cohesion
Operational Programme (Grant KK.01.1.1.01.0004).
The  work of LMR was 
supported by Spanish Ministry of Economy and Competitiveness (MINECO)
Grants No. FPA2015-65929-MINECO and FIS2015-63770-MINECO.

\bibliography{refs}

\end{document}